\documentstyle[12pt]{article}
\renewcommand{\theequation}{\thesection.\arabic{equation}}
\newlength{\extraspace}
\newlength{\extraspaces}
\newcommand{\be}{\begin{equation}
\addtolength{\abovedisplayskip}{\extraspaces}
\addtolength{\belowdisplayskip}{\extraspaces}
\addtolength{\abovedisplayshortskip}{\extraspace}
\addtolength{\belowdisplayshortskip}{\extraspace}}
\newcommand{\ee}{\end{equation}}
 
\newcommand{\ba}{\begin{eqnarray}
\addtolength{\abovedisplayskip}{\extraspaces}
\addtolength{\belowdisplayskip}{\extraspaces}
\addtolength{\abovedisplayshortskip}{\extraspace}
\addtolength{\belowdisplayshortskip}{\extraspace}}
\newcommand{\ea}{\end{eqnarray}}

\newcommand{\bas}{\begin{eqnarray*}
\addtolength{\abovedisplayskip}{\extraspaces}
\addtolength{\belowdisplayskip}{\extraspaces}
\addtolength{\abovedisplayshortskip}{\extraspace}
\addtolength{\belowdisplayshortskip}{\extraspace}}
\newcommand{\eas}{\end{eqnarray*}}
 
\newcounter{subequation}[equation]
\makeatletter

\expandafter\let\expandafter
\reset@font\csname reset@font\endcsname

\def\subeqnarray{\arraycolsep1pt
    \def\@eqnnum\stepcounter##1{\stepcounter{subequation}%
        {\reset@font\rm(\theequation\alph{subequation})}}
\jot5mm     \eqnarray}

\makeatother

\newcommand{\newappendix}[1]{
\vspace{15mm}
\pagebreak[3]
\addtocounter{section}{1}
\setcounter{equation}{0}
\setcounter{subsection}{0}
\setcounter{footnote}{0}
\renewcommand{\theequation}{\Alph{section}.\arabic{equation}}
\begin{flushleft}
{\large\bf Appendix \Alph{section}: #1}
\end{flushleft}
\nopagebreak
\medskip
\nopagebreak}

\def\eql#1{\\(#1)\vadjust{\penalty10000\vskip-5.3ex}}  
\def\eqlh#1{\\(#1)\vadjust{\penalty10000\vskip-5.5ex}}  

\newcommand{\newsection}[1]{
\vspace{15mm}
\pagebreak[3]
\addtocounter{section}{1}
\setcounter{equation}{0}
\setcounter{subsection}{0}
\setcounter{footnote}{0}
 
\begin{flushleft}
{\large\bf \thesection. #1}
\end{flushleft}
\nopagebreak
\medskip
\nopagebreak}
 
\newcommand{\newsubsection}[1]{
\vspace{1cm}
\pagebreak[3]
 
\addtocounter{subsection}{1}
\noindent{ \bf \thesection.\arabic{subsection} #1}
\nopagebreak
\vspace{2mm}
\nopagebreak}

\newcommand{\NP}[1]{Nucl.\ Phys.\ {\bf #1}}

\newcommand{\CMP}[1]{Comm.\ Math.\ Phys.\ {\bf #1}}

\newcommand{\N}{\mbox{I\hspace{-.4ex}N}}
\newcommand{\C}{\mbox{$\,${\sf I}\hspace{-1.2ex}{\bf C}}}
\newcommand{\Z}{\mbox{{\sf Z}\hspace{-1ex}{\sf Z}}}
\newcommand{\R}{\mbox{\rm I\hspace{-.4ex}R}}
\newcommand{\1}{\mbox{1\hspace{-.6ex}1}}
\newcommand{\bra}{\langle}
\newcommand{\ket}{\rangle}
\newcommand{\ra}{\rightarrow}

\newcommand{\rra}{\ \longrightarrow \ }

\newcommand{\is}{ &\! =\! & }
\newcommand{\nonum}{\nonumber \\[1.5mm]}
\newcommand{\sspace}{\makebox[1cm]{ }}
\newcommand{\bspace}{\makebox[2cm]{ }}
\newcommand{\nspace}{\!\!\!\!\!\!\!\!\!\!}

\newcommand{\inv}{^{-1}}

\newcommand{\th}{{\theta}}
\newcommand{\lb}{\lambda}

\newcommand{\sbar}{{\overline{s}}}

\newcommand{\Dbar}{{\overline{D}}}

\newcommand{\Gbar}{{\overline{G}}}
\newcommand{\Hbar}{{\overline{H}}}
\newcommand{\Tbar}{{\overline{T}}}

\newcommand{\Wbar}{{\overline{W}}}

\newcommand{\cA}{{\cal A}}
\newcommand{\cB}{{\cal B}}

\newcommand{\cF}{{\cal F}}

\newcommand{\cH}{{\cal H}}

\newcommand{\cN}{{\cal N}}
\newcommand{\cM}{{\cal M}}
\newcommand{\cO}{{\cal O}}

\newcommand{\cT}{{\cal T}}
\newcommand{\cW}{{\cal W}}

\newcommand{\ob}{\omega_{\beta}}

\newcommand{\cbM}{\overline{\cal M}}
\newcommand{\cbN}{\overline{\cal N}}
\newcommand{\cbF}{\overline{\cal F}}
\newcommand{\Obra}{{\langle \Theta_{\beta}|}}
\newcommand{\Oket}{{|\Omega_{\beta}\rangle}}

\begin{document}
%
\begin{titlepage}
%
\renewcommand{\thefootnote}{\fnsymbol{footnote}}
\mbox{ }
\vspace{15mm}

\begin{center}
{\LARGE Form Factors, Thermal States and Modular Structures}\\[2.5cm]
{\large Max R. Niedermaier}\\ [3mm]
{\small\sl Max-Planck-Institut f\"{u}r Gravitationsphysik} \\
{\small\sl (Albert-Einstein-Institut)} \\
{\small\sl Schlaatzweg 1, D-14473 Potsdam, Germany}
\vspace{3.5cm}
 
{\bf Abstract}
\end{center}
\begin{quote}
Form factor sequences of an integrable QFT can be defined axiomatically 
as solutions of a system of recursive functional equations, known as 
``form factor equations''. 
We show that their solution can be replaced with the study of the 
representation theory of a novel algebra $\cF(S)$. It is associated 
with a given two-particle S-matrix and has the following features:
(i) It contains a double TTS algebra as a subalgebra.
(ii) Form factors arise as thermal vector states over $\cF(S)$ of 
temperature $1/2\pi$. The thermal ground states are in correspondence 
to the local operators of the QFT.
(iii) The underlying `finite temperature structure' is indirectly 
related to the ``Unruh effect'' in Rindler spacetime. In $\cF(S)$ it is 
manifest through modular structures $(j,\delta)$ in the sense of 
algebraic QFT, which can be implemented explicitly in terms of the 
TTS generators.
\end{quote}
\vfill

\renewcommand{\thefootnote}{\arabic{footnote}}
\setcounter{footnote}{0}
\end{titlepage}
\newsection{Introduction and survey}

The form factor approach to integrable quantum field theories
(QFTs) has several remarkable features. First it yields 
a complete (`non-perturbative') definition of an integrable
QFT, independent of any Lagrangian description. Rather  
the QFT is described in terms of ``form factors'', which arise as 
elements of sequences of tensor-valued meromorphic functions, solving a 
recursive set of functional equations. Second it provides a powerful 
non-perturbative solution technique, yielding results difficult or 
impossible to obtain otherwise. This is especially relevant when
the QFT in question has an independent description as, say, the 
continuum limit of some lattice system. 
Third, no regularization and renormalization is necessary to
obtain genuine QFT quantities. From the viewpoint of Haag's 
theorem the last property hints at the existence of an 
underlying algebraic structure with a controllable 
representation theory, in the sense that the representations
relevant for the specification of an interacting QFT can be 
described explicitly (unlike the situation for the canonical
commutation relations).

\newsubsection{Form factor sequences in integrable QFTs}

The algebraic analysis of form factors in an integrable QFT is 
usually done in terms of the Zamolodchikov-Faddeev (ZF) 
algebra. Originally the ZF-algebra was invented to give a concise
description of all the S-matrix elements of an integrable QFT
\cite{ZZ,Fadd}. Motivated by the fact that also form factors
obey S-matrix exchange relations they have been interpreted as 
linear functionals over the ZF algebra \cite{Smir,Luk1}. 
Here we shall argue that form factors should more appropriately be 
regarded as functionals over a larger algebra \cite{MNalg}.
In upshot each form factor sequence turns out to be in 
correspondence to a linear functional
over a novel algebra $\cF_{\beta}(S)$, which we call ``form factor 
algebra''. The relevant functionals are characterized by a simple 
invariance condition (``T-invariance''). From the quantum field 
theoretical perspective, for $\beta =2\pi$, each form factor sequence 
corresponds to a local operator in the QFT aimed at. 
Symbolically therefore one has the correspondences:
$$
\begin{tabular}{c} 
Local Operator \\ in QFT 
\end{tabular}
\;\;\longleftrightarrow\;\; 
\begin{tabular}{c} 
Form Factor \\ Sequence  
\end{tabular}
\;\;\longleftrightarrow\;\; 
\begin{tabular}{c} 
T-invariant Functional\\
over $\cF_{2\pi}(S)$  
\end{tabular}
$$

When $\beta$ is different from $2\pi$ the corresponding sequences 
or functionals turn out to deformed QFTs with still having the 
same bootstrap S-matrix as the original QFT \cite{MNdeform}.

The motivation to develop such an algebraic framework is three-fold.
First the formulation avoids making reference to ill-defined traces
inevitably showing up otherwise (c.f. below). Second it is amenable 
to generalizations not visible in the QFT context. Examples are
the deformed QFTs mentioned before and applications to 
the quantum Ernst equations. Third, one may hope it to lead to more
useful expressions for the form factors in concrete models.   
So far mostly the solutions of the form factor equations are obtained 
in the form of multiple contour integrals, where the 
integrand has complicated monodromy properties \cite{Smir1,BBS}. 
Since for example
a two-point function gets computed from multiple integrals of
the modulus square of a form factor, such expressions -- 
though mathematically intriguing in its own right  -- are
not of immediate practical use. Of course it remains to be seen whether
the framework here can improve on this.

\newsubsection{Finite temperature structure}

An important feature of the algebra $\cF_{2\pi}(S)$ is that it 
unravels the `finite temperature structure' underlying the form 
factor approach. The point at issue can be seen from the cyclic form 
factor equation stating that  
\be
F_{a_n\ldots a_1}(\th_n+ 2\pi i, \ldots ,\th_2,\th_1) =
\eta\,F_{a_{n-1}\ldots a_1 a_n}(\th_{n-1},\ldots,\th_1,\th_n)\;,
\label{I1}
\ee
where $\eta$ is a phase and the shift by $2\pi i$ is understood in 
the sense of analytic continuation. Originally equation (\ref{I1}) 
was found in the context of the Sine-Gordon model
\cite{SmirSG} improving on earlier attempts to generalize Watsons 
equation \cite{KarWeisz}. Subsequently Smirnov promoted it to an axiom
for the form factors of an integrable QFT, which together with the 
other equations implies locality \cite{Smir}.
More recently a derivation of (\ref{I1}) from quantum field theoretical 
principles was given, showing in particular that (\ref{I1}) holds in any 
massive 1+1 dim. relativistic QFT, regardless of its integrability 
\cite{MNcycl}. The origin of the phase $\eta$ is clarified in \cite{Rehren}.

Evidently equation (\ref{I1}) has the form of a thermal state, or 
Kubo-Martin-Schwinger (KMS) condition, with the parameter of the Lorentz 
boosts playing the role of a `time' variable. This immediately 
suggests a relation to the Unruh effect \cite{Luk2,Schroer,MNunpubl}, 
which however is not straightforward
to unravel in the formalism: The Unruh effect proper is strictly 
limited to free QFTs, in that the Bogoliubov transformation employed
in its derivation cannot be constructed otherwise. Its generalization 
to interacting QFTs is given by the Bisognano-Wichmann theorem \cite{BW}. 
Both lead to KMS conditions for Wightman functions in (complexified) 
position space. Equation (\ref{I1}) can of course {\em not} be 
obtained by simply Fourier transforming such a position space KMS 
condition (as sometimes asserted). Rather equation (\ref{I1}) is a 
statement about matrix elements of scattering states, not about Wightman 
functions. In particular all momenta in (\ref{I1}) are on-shell momenta 
which arise through the asymptotic clustering in a Haag-Ruelle type 
construction of the scattering states. The thermal properties then
have to be rederived from scratch. More details on the derivation 
of (\ref{I1}) can be found in \cite{MNcycl}.

In the integrable case, after switching to an appropriate 
pseudo particle basis, the form factors will in addition obey the 
familiar S-matrix exchange relations. Being a KMS condition 
(\ref{I1}) then superficially suggests to search for 
solutions of (\ref{I1}) in terms of traces over ZF-type operators
\be
F_{a_n\ldots a_1}(\th_n, \ldots ,\th_1) 
\stackrel{\displaystyle ?}{=} Tr\left[e^{2\pi K}\cO\,
V_{a_n}(\th_n)\ldots V_{a_1}(\th_1)\right]\;,
\label{I2}
\ee 
where $K$ is the generator of Lorentz boosts, $\cO$ represents the local
operator and 
$V_a(\th)$ are ZF-type operators satisfying
\be
V_a(\th_1)\, V_b(\th_2) = S^{dc}_{ab}(\th_{21})\;
V_c(\th_2)\, V_d(\th_1)\;,\;\;\;{\rm Re}\,\th_{21} \neq 0\;,\;\;
\th_{21}=\th_2 -\th_1\,.
\label{I3}
\ee
Indeed, such a construction works nicely in the context of lattice 
models \cite{JimMiwa1} but in a QFT context the relevant trace will never 
exist and 
for a very physical reason. Namely $K$ is {\em not} like a Hamiltonian, 
its spectrum is not bounded from below but consists of the entire real axis. 
In other words the `density operator' $e^{2\pi K}$ is an unbounded 
operator (and so will be in general also the ZF operators) and the
trace is meaningless on any state space on which $K$ has the proper 
spectrum. Of course one can try to disregard this as a technical nuisance 
and `renormalize' the trace in various ways. For example one can divide 
(\ref{I2}) by the equally divergent $Tr[e^{2\pi K}]$ \cite{KLP2,EfLeCl,Luk2}.
In general however there will be no guarantee that the ratio is finite; 
for example the divergence may depend on $\th_n,\ldots,\th_1$ or on the 
local operator considered. One can also return to a lattice formulation
and try to find the proper thermodynamic limit \cite{Lash}, though
the trace interpretation of (\ref{I2}) is unlikely to survive the limit 
\cite{HHW}. Any such procedure however requires a regularization that 
spoils some of the fundamental features of the QFT aimed at. In addition 
it is model dependent and against the spirit of the form factor 
approach, whose most compelling feature is that no regularization and no 
renormalization is necessary to construct genuine QFT quantities.

In the spirit of the form factor approach one can ask whether 
it is possible to replace the inevitably sick trace in (\ref{I2}) 
systematically by something well-defined. Clearly the required 
mathematics must be able to deal with thermal states having an 
unbounded density operator. Fortunately the relevant mathematics is known 
for almost 30 years and has been found independently in the context of QFT 
at finite temperature \cite{HHW} and the structure analysis of von Neumann 
algebras \cite{TT}. The by now common heading is: {\bf Modular Structures}. 
It may be helpful to briefly recapitulate their basic features. 
Modular structures in the context of von Neumann algebras are a 
pair of operators $(J,\Delta)$ that can be associated to any 
von Neumann algebra $\cN$ with cyclic and separating vector $\Omega$.
The latter means that there exists a Hilbert space $\cH$ 
such that both $\cN\Omega$ and $\cN'\Omega$ are dense subspaces 
of $\cH$, where $\cN'$ is the commutant of $\cN$, that is the set 
(and $C^*$-algebra) of all bounded operators on $\cH$ commuting with $\cN$.
The operator $J$ is an antiunitary involution with respect to the inner 
product on $\cH$, and $\Delta$ is a positive selfadjoint (in general 
unbounded) operator. Further they obey the following defining relations
\ba
J \Delta^{1/2} X \Omega \is X^* \Omega \;,\;\;\;\;\mbox{for}\;X \in \cN\;,
\nonum
J \Delta^{-1/2} X' \Omega \is {X'}^* \Omega\;,\;\;\;\mbox{for}\; 
X' \in \cN'\;,\;\;\; \mbox{with}\;\;\; J\Delta J = \Delta^{-1}. 
\label{I4}
\ea
The Tomita-Takesaki theorem \cite{TT} states that
$J \cN J = \cN'$ and that for all real $\lb$ the mapping
$D_{\lb}(X) = \Delta^{i\lb} X \Delta^{-i\lb}$ defines an automorphism 
group of both $\cN$ and $\cN'$. In this context we shall refer to the 
following equation as the ``KMS property'' of $\Delta$
\be
(\Omega, Y \Delta X\Omega) = (\Omega, X Y\Omega)\;,\;\;\;X,Y \in \cN\;.
\label{I5}
\ee
It follows from the defining relations via \cite{HHW,TT}
\ba
&& (\Omega, Y \Delta X\Omega) = (\Delta^{1/2} Y^*\Omega\,,
\Delta^{1/2} X \Omega) \nonum 
&& =(J Y \Omega\,,J X^*\Omega) =
(X^*\Omega\,,Y\Omega) = (\Omega, X Y\Omega)\;.
\label{I6}
\ea
Heuristically one can think of $\Delta$ as being an unbounded 
density operator for which the relations (\ref{I4}), (\ref{I5}) 
provide a substitute for the cyclic property of the trace.

In the context of form factors one is not naturally 
dealing with von Neumann algebras and the above results do not apply. 
Nevertheless one can try to develop an algebraic counterpart
of this construction, just as the ``thermofield formalism'' 
\cite{Umz} takes the above algebraic relations as the 
starting point \cite{Ojima}, ignoring topological issues in practice.
At least for the time being such an algebraic viewpoint seems 
to be appropriate. Transferred to the form factor situation, a 
counterpart of the `modular' or `thermofield' formalism  can be 
formulated as follows. First one gives up the presupposition that the 
algebra (\ref{I3}) is represented irreducibly on the state space. 
Rather one works with a manifestly reducible representation.
If $W_a(\th)$ denotes the generators of (\ref{I3}) in this representation,
the reducibility is manifest in that there exists a large class of 
operators commuting with the $W_a(\th)$'s, namely a set of operators
$j(W_a)(\th)$ satisfying (\ref{I3}) with the complex conjugate S-matrix 
(the ``tilde fields'' in the thermofield language). Thus  
\ba
W_a(\th_1)\, W_b(\th_2) \is S^{dc}_{ab}(\th_{21})\;
W_c(\th_2)\, W_d(\th_1)\;,
\nonum
j(W_a)(\th_1)\, j(W_b)(\th_2) \is [S^{dc}_{ab}(\th_{21})]^*\;
j(W_c)(\th_2)\, j(W_d)(\th_1)\;,\;\;\; {\rm Re}\,\th_{21} \neq 0\;.
\nonum
W_a(\th_1)\, j(W_b)(\th_2) \is j(W_b)(\th_2)\,W_a(\th_1)\;.
\label{I7}
\ea
For real boost parameters $\lb$ the Lorentz boosts act as 
automorphisms of the algebra (\ref{I7}) via $D_{\lb}(W_a)(\th) = 
e^{i\lb K} W_a(\th) e^{-i\lb K} = W_a(\th + \lb)$, etc..
For imaginary boost parameters $\lb$ this is simply meaningless. 
One of the main results obtained here is as follows: If one 
starts with an algebra containing a double TTS algebra in addition 
to an algebra of the form (\ref{I3}) and imposes one extra relation, 
then both $j(W_a)(\th)$ and $D_{2\pi i}(W_a)(\th)$ can be expressed 
{\em explicitly} in terms of $W_a(\th)$ and the generators 
$T^{\pm}(\th)_a^b$ of the double TTS algebra. Namely
\ba
j(W_a)(\th) \is C_{aa'} C^{mn}W_m(\th^*)\,
T^+(\th^* +i\pi)_n^{a'} \nonum
\is C_{aa'}C^{mn}T^-(\th^* +i\pi)_n^{a'}W_m(\th^* + i2\pi)\;, 
\nonum 
D_{2\pi i}(W_a)(\th) \is C_{mn}T^-(\th +i2\pi)_a^m \,W_k(\th)\,
T^+(\th +i\pi)_l^n C^{lk} = W_a(\th + i2\pi)\;.
\label{I8}
\ea
Here it is stipulated that $D_{2\pi i}(W_a)(\th)=
W_a(\th + i2\pi)$ is a relation in the new algebra, which we 
call ``modular'' algebra $\cM_{2\pi}(S)$, where S refers to 
the given S-matrix and $2\pi$ is the inverse temperature  
featuring in the KMS condition. The term ``modular'' is used 
because the assignments $j$ and $D_{2 \pi i}$ turn out to be 
(anti-linear and linear) automorphisms of $\cM_{2\pi}(S)$ having 
all algebraic properties  of modular operators in the context of 
von Neumann algebras, in particular $j^2 = id$.  
The counterpart of the ``thermal ground state'' (the cyclic vector 
in the context of von Neumann algebras) are vectors 
$|\Omega_{2\pi}\ket$ and $\bra \Theta_{2\pi}|$ satisfying 
\be
T^+(\th)_a^b|\Omega_{2\pi}\ket = \delta_a^b |\Omega_{2\pi}\ket \;,\;\;\;  
\bra\Theta_{2\pi}|T^-(\th)_a^b = \eta \,\delta_a^b \bra\Theta_{2\pi}|\;,
\label{I9}
\ee   
where in general $\bra\Theta_{2\pi}|\Omega_{2\pi}\ket =0$. One of these 
vectors, say $\bra\Theta_{2\pi}|$, should eventually be 
thought of as being in correspondence to a local operator in the 
QFT considered, the other can be viewed as a version of the 
``Rindler vacuum''. The matrix elements
\be
F_{a_n\ldots a_1}(\theta_n,\ldots,\theta_1)
=\langle \Theta_{2\pi}| W_{a_n}(\theta_n)\ldots W_{a_1}(\theta_1)  
|\Omega_{2\pi} \rangle \;
\label{I10}
\ee
then automatically satisfy (\ref{I1}). The usefulness of a ``Yangian'' 
or ``quantum double'' extension of the ZF-algebra has been noticed by 
a number of authors \cite{DeVega1,DeVega2,Smir3,BeLeCl,KLP,Ding}, however 
without employing the crucial relation (\ref{I8}), denoted by (M) (for
``modular'') below. To simplify the discussion we 
ignored the residue condition so far. By a suitable modification 
$\cF_{2\pi}(S)$ of the algebra $\cM_{2\pi}(S)$ one can achieve    
that the matrix elements (\ref{I7}) in fact satisfy all the 
form factor equations of an integrable QFT without bound states. 
Correspondingly we refer to $\cF_{2\pi}(S)$ as the ``form factor algebra''.
The solution of the recursive system of form factor equations can 
thus be replaced with the study of the representation 
theory of $\cF_{2\pi}(S)$. 

The rest of the paper is organized as follows. In the next section
we define the algebra $\cF_{\beta}(S)$, allowing for 
$\beta$ different from $2\pi$, and establish the (right half of the) 
correspondence displayed in section 1.1. The emergence of the modular 
structures $(j,\delta)$ is described in section 3, to be followed 
by a brief outlook on the perspective.


\newsection{An algebra underlying the form factor equations}

We work with a slightly generalized set of form factor equations, 
depending on a real parameter $\beta$. A detailed description is relegated 
to appendix A. For $\beta =2\pi$ they coincide with the form factor 
equations of an integrable massive QFT without bound states. For 
generic $\beta$ one obtains a system of deformed form factor equations, 
whose solutions turn out to define QFTs with a deformed kinematical
arena having the same bootstrap S-matrix as the original QFT 
\cite{MNdeform}.  
Conceptually the solutions to both systems of equations are sequences 
of tensor-valued meromorphic functions. Here we show that 
such sequences can be set into correspondence to linear functionals
over an algebra $\cF_{\beta}(S)$, which we call ``form factor 
algebra''. The relevant functionals are characterized by 
the ``T-invariance'' condition (\ref{I9}).  

The algebra $\cF_{\beta}(S)$ will defined mainly in terms of 
(quadratic) relations among its generators. We deliberately
refrain from introducing topological notions here for two 
reasons. First, the appropriate topology is better specified 
together with and in terms of the T-invariant functionals 
mentioned before. Secondly, even for the much simpler case 
of the Yangian double the rigorous reconstruction as a 
kind of ``braided vertex operators algebra'' \cite{EtKa} or 
``deformed chiral algebra'' \cite{FrRe} has only begun and 
rests on the specific form of the S-matrix. The first part
of the following definition therefore is somewhat schematic 
but sufficient for the purposes here.  

\newsubsection{Definition of the algebra}

Let $\cA$ denote an abstract normed $*$-algebra equipped with an 
$\N$-grading such that the degrees add up upon multiplication of 
two elements. Let $\cA^{(n)}$ be the subspace of degree n and
consider the space of mappings $\cA^{(m,n)}:D^m \ra \cA^{(n)}$  
continuous on a subset $D^m= (\R + I\Z)^m$ of $\C^m$, where $I$ is 
a finite set of purely imaginary numbers.
One expects that such mappings can be generated by
`suitable' multiplication of elements of $\cA^{(1,1)}$, i.e. of 
1-parameter families $\th \ra X(\th)$ of degree 1 operators.
However, in general the product of two elements of $\cA^{(1,1)}$ will not
be continuous on $D^1\times D^1$. We call 
\be 
\cA({\rm data}) = \bigoplus_{m \geq n} \cA^{(m,n)}
\label{qop1}
\ee
a ``braided vertex operator algebra'' (braided VOA)%
\footnote{We borrow the term, though not the concept from \cite{EtKa}.}
if the elements of $\cA^{(m,n)}$
are generated by two types of product operations from $\cA^{(1,1)}$
(and indicate in brackets what kind of data the construction depends on).
First, the ordinary product, which will be denoted simply by 
concatenation of generators. It is assumed to be well-defined (at least)
whenever ${\rm Re}\,\th_1 \neq {\rm Re}\,\th_2$ for $X_1(\th_1),\,X_2(\th_2) 
\in \cA^{(1,1)}$. By iteration elements of $\cA^{(n,n)}$ for arbitrary $n$ 
can be generated and will be of the form $X_1(\th_1) \ldots X_n(\th_n)$ 
with ${\rm Re}\,\th_i \neq {\rm Re}\,\th_j,\;i\neq j$. Associativity of this
product is a consequence of the associativity of the underlying 
algebra $\cA$. In general not all elements of $\cA^{(n,n)}$ will be 
linearly independent. The linear dependencies are induced by   
exchange relations to be specified later involving the ``braiding matrix''. 
Second we assume that there exists a contraction product, 
which is defined whenever a difference $\pm(\th_1-\th_2)$ 
assumes one out of the finite number of purely imaginary values $I$.
The contraction product is again defined recursively starting from  
\be
\cdot \;: \cA^{(1,1)} \times \cA^{(1,1)} \rra  \cA^{(1,0)}\;.
\label{qop2}
\ee
The extension to other $\cA^{(m,n)}$ is done by assuming compatibility
with the ordinary product $X\cdot(YZ) = (X\cdot Y)Z$. In general the 
contraction product is neither commutative nor associative.
For multiple contraction products we use a right nesting convention, 
i.e. we write $X\cdot Y \cdot Z$ for $X\cdot(Y\cdot Z)$ etc.  
Clearly a braided VOA can be specified in terms of 
the generators $\cA^{(1,1)}$ and $\cA^{(1,0)}$, the set $I$, and the 
two products. Note that elements of infinite degree are not defined.
A braided VOA will be called {\em Lorentz covariant} if translations 
in the variable ${\rm Re}\,\th$ are 
unitarily implemented; the generator of the automorphism group will be 
denoted by $K$. Explicitly on the degree 1 elements this means 
$e^{i\lb K} X(\th) e^{-i\lb K} = X(\th + \lb),\;\lb \in \R$. 

In the following we define three Lorentz covariant braided VOAs 
associated with a given bootstrap S-matrix.
In addition they depend on a real parameter $\beta$ and the  
index set $I = \{i\pi,i(\beta -\pi)\}$. The charge conjugation   
matrix and its inverse are identified with central elements 
$C_{ab},\,C^{ab}\in \cA^{(0,0)}\subset \cA^{(1,0)}$, where the 
inclusion treats $C_{ab}, C^{ab}$ as constant functions in $\cA^{(1,0)}$.
We shall write $\cA_{\beta}(S)$ for $\cA(S,I,\beta)$ in (\ref{qop1})
and use different symbols for the various algebras, but keep 
the generic notation $\cA^{(m,n)}$ for their grade spaces.
The algebras are:
\begin{enumerate}
\item A generalized {\em quantum double} $\cT_{\beta}(S)$ with 
generators $T^{\pm}(\th)_a^b$.
\item The {\em modular algebra} $\cM_{\beta}(S)$ with generators 
$T^{\pm}(\th)_a^b$ and $W_a(\th)$.
\item The {\em form factor algebra} $\cF_{\beta}(S)$ with the same 
generators as $\cM_{\beta}(S)$ but extra relations.
\end{enumerate}
The indices $a,b$ etc.~refer to the modules $V_a,V_b$ associated with 
the given bootstrap S-matrix $S$, c.f appendix A. The grading is such 
that $W_a(\th),\;T^{\pm}(\th)_a^b\in \cA^{(1,1)}$.
The two products will be defined by specifying relations among the 
generators.

\begin{center}
\fbox{Definition of $\cT_{\beta}(S)$ and $\cM_{\beta}(S)$}
\end{center}

The defining relations of $\cT_{\beta}(S)$ are:
\vspace{3mm}
\eql{T1}
\bas \jot5mm
&& S^{cd}_{mn}(\th_{12})\,T^{\pm}(\th_1)_a^nT^{\pm}(\th_2)_b^m
= T^{\pm}(\th_2)_n^c T^{\pm}(\th_1)_m^d\,S_{ab}^{mn}(\th_{12})\;,\nonum
&& S^{cd}_{mn}(\th_{12})\,T^+(\th_1)_a^nT^-(\th_2)_b^m =
T^-(\th_2)_n^c T^+(\th_1)_m^d\,S_{ab}^{mn}(\th_{12}+i 2\pi-i\beta)\;,
\eas
valid for $Re\,\th_{12}\neq 0$, $\th_{12}:= \th_1 -\th_2$. Further
\vspace{3mm}
\eql{T2}
\bas \jot5mm
&& C_{mn}T^{\pm}(\th)_a^m\cdot T^{\pm}(\th-i\pi)_b^n=C_{ab}\;,\nonum
&& C^{mn}T^{\pm}(\th)_m^a \cdot T^{\pm}(\th+i\pi)_n^b=C^{ab}\;.
\eas
The `$\;\cdot\;$' product on $\cT_{\beta}(S)$ is 
associative. 

Now we extend the algebra $\cT_{\beta}(S)$ to $\cM_{\beta}(S)$ by adding
generators $W_a(\th)$ having the following 
linear exchange relations with $T^{\pm}(\th)_a^b$\hfill
\vspace{4mm}
\eql{TW}
\bas\jot5mm
&& T^-(\th_1)_a^e\,W_b(\th_2)=S^{dc}_{ab}(\th_{12})\,
W_c(\th_2)\,T^-(\th_1)_d^e\;,\nonum
&& T^+(\th_1)_a^e\,W_b(\th_2)=S^{dc}_{ab}(\th_{12}+i 2\pi-i\beta)\,
W_c(\th_2)\,T^+(\th_1)_d^e\;. 
\eas
These relations hold for all relative rapidities, including 
${\rm Re}\,\th_{12} = 0$. Further we impose
\vspace{4mm}
\eql{WW}
\bas \jot5mm
W_a(\th_1)\,W_b(\th_2) \is S^{dc}_{ab}(\th_{12})\;
W_c(\th_2)\,W_d(\th_1) \;,\sspace Re\,\th_{12}\neq 0\;.
\eas
\vspace{-3mm}
\eql{M}
\bas\jot5mm
&& C^{mn}\,W_m(\th)\cdot T^+(\th +i\beta -i\pi)_n^a = 
C^{mn}\,T^-(\th +i\beta -i\pi)_n^a \cdot W_m(\th +i\beta)\;.
\eas
The relation (M) will later turn out to implement the action of 
the modular operators on $\cM_{\beta}(S)$. The `$\;\cdot\;$' product
on $\cM_{\beta}(S)$ is associative. An equivalent form of (M) is 
\ba
&& W_a(\th +i\beta) = C_{mn}T^-(\th +i\beta)_a^m \cdot W_k(\th)
                 \cdot T^+(\th+i\beta-i\pi)_l^n\,C^{kl}\;,\nonum
&& W_a(\th -i\beta) = C^{kl}T^-(\th -i\pi)_l^n \cdot W_k(\th) \cdot 
                    T^+(\th-i2\pi)_a^m\,C_{mn}\;.
\label{w1}
\ea

\begin{center}
\fbox{Definition of $\cF_{\beta}(S)$}
\end{center}

Now we supplement the $\cM_{\beta}(S)$ algebra by an extra relations 
for the contraction products of $W$-generators and denote the resulting
algebra by  $\cF_{\beta}(S)$.
\vspace{4mm}
\eql{R}
\bas
\nspace\mbox{$\beta$ generic:}
&& W_a(\th + i\pi)\cdot W_b(\th) = -\lb\,C_{ab}\;,\nonum
&& C^{ab}W_a(\th - i\pi)\cdot W_b(\th) = - \lb\;,\nonum
\nspace \mbox{$\beta=2\pi$:}
&& W_a(\th +i\pi)\cdot W_b(\th) = \lb\,[D^+_{ab}(\th)- \,C_{ab}]\;,\nonum
&& C^{ab}W_a(\th -i\pi)\cdot W_b(\th) = \frac{\lb}{\dim V}\,C^{ab}
[D^-_{ab}(\th-i\pi) - C_{ab}]\;,
\eas
where $\lb \in \R$ and we set 
\ba
&& D^+_{ab}(\th) := C_{mn}\,T^-(\th+i\pi)_a^m \cdot T^+(\th)_b^n\;,\nonum
&& D^-_{ab}(\th) := - \dim V \,S_{ab}^{dc}(i\pi-i\beta)
D^+_{cd}(\th)\;.
\label{D1}
\ea
For a $2\pi i$-periodic S-matrix the $C^{ab}$ contraction in the 
$W_a(\th - i\pi)\cdot W_b(\th)$ product can be dropped. 
For $\beta =2\pi$ we require in addition that $0 \leq {\rm Im}\,\th
\leq \pi$ in (R); the extension to other strips is done by means of (M).
In particular using (M) and (TW) the second $\beta =2\pi$ equation 
can be seen to be a consequence of the first one. Note also that 
for $\beta =2\pi$ the contraction $C^{ab}D^+_{ab}(\th)=
C^{ab}D^-_{ab}(\th)$ is defined even when $S_{ab}^{cd}(-i\pi)$ is 
singular, because $-\dim V S_{ab}^{cd}(-i\pi) C_{cd} = C_{ab}$ is 
always regular. This concludes the definition of the algebras
1.--3.  

Implicit in these definitions, of course, is the presupposition 
that the above relations define consistent algebras. 
\medskip

{\bf Proposition:} The algebras 1.--3. are consistent (in the sense 
of the proof). The ordinary product is associative and compatible 
with the contraction product.
\medskip

Proof. (Sketch) One has to show that the defining relations for both
the ordinary product and the contraction product arise from dividing
out two-sided ideals in the respective `free' algebras, where no 
relations  among the generators are imposed. Once this is known, 
associativity of the ordinary product follows from the assumed 
associativity of the underlying $*$-algebra. Consistency with 
the contraction product is assumed when checking the ideals and then
justified in retrospect. The search for ideals and the construction
of the successive quotient algebras is best done in a particular
order. The principle is largely analogous to that in \cite{MNalg}, 
so that it may be sufficient to list the items to be checked 
and to make a few comments on each entry:
\begin{itemize}
\item[(a)] Consistency and associativity of $\cT_{\beta}(S)$: 
In lack of a centrally extended quantum double construction we 
check the consistency and associativity directly. 
For relations (T1) there are in principle 
six consistency conditions to be checked, which arise from pushing 
$T^+$ or $T^-$ through one of the relations (T1). Using the known 
consistency of the $\beta$-independent equations and the homomorphism 
(\ref{iso1}) below, only two of them have to be checked 
explicitly. Doing this one establishes the consistency and, as a 
byproduct, the associativity of the algebra with relations (T1).
In this algebra the relations (T2) are found to correspond 
to two-sided ideals. Dividing out these ideals one establishes the 
consistency of $\cT_{\beta}(S)$ and its associativity with respect to both 
products. 
\item[(b)] Consistency of (TW) and (WW) with $\cT_{\beta}(S)$: 
First one checks the consistency of (TW) with $\cT_{\beta}(S)$ by 
pushing $T^{\pm}$ through (TW) and $W$ through (T1), (T2). Similarly
one verifies the consistency of (TW) and (WW). Finally one shows that
the relations (T1), (T2) arise from two-sided ideals in the 
associative algebra with relations (TW) and (WW) only. 
\item[(c)] Consistency of (M) with all other relations: 
One can verify that (M) arises from a two-sided ideal in the algebra 
with relations (TW), (WW), (T1). This fact holds for any  
relative coefficient between the left and the right hand side of (M).
In particular the phase $\eta$ appearing in (\ref{I9}) could also be 
incorporated here. It is however more natural to attribute $\eta$ to 
the state rather than the algebra. Dividing out the ideal corresponding
to (M) the consistency of the algebra $\cM_{\beta}(S)$ follows.
\item[(d)] Consistency of (R) with all other relations: Again both 
sides of the relations (R) have to generate two-sided ideals in the 
algebra without the relations imposed. For $\beta$ generic this
is straightforwardly verified. For $\beta = 2\pi$ it is a consequence 
of the following exchange relations between $D^+_{ab}(\th)$ and 
$T^{\pm}(\th)_a^b\;,W_a(\th)$:
\ba
S_{ab}^{mn}(\th_{21}+i\beta -i2\pi)\,D^+_{cn}(\th_1)\,T^-(\th_2)_m^d
\is S_{ac}^{mn}(\th_{12}+i\pi)\,T^-(\th_2)_m^d\,D^+_{nb}(\th_1)\;,
\nonum
S_{ab}^{mn}(\th_{21})\,D^+_{cn}(\th_1)\,T^+(\th_2)_m^d
\is S_{ac}^{mn}(\th_{12}-i\pi + i\beta)\,T^+(\th_2)_m^d\,
D^+_{nb}(\th_1)\;\nonum
S_{ab}^{mn}(\th_{21}+i\beta -i2\pi)\,D^+_{cn}(\th_1)\,W_m(\th_2)
\is S_{ac}^{mn}(\th_{12}+i\pi)\,W_m(\th_2)\,D^+_{nb}(\th_1)\;.
\label{c3}
\ea
\end{itemize}
This concludes the verification of the consistency of the algebras 
1.--3.. Next we discuss some of their basic properties. 

\newsubsection{Basic properties}

The algebra $\cT_{\beta}(S)$ is a 
well-known structure. For $\beta = 2\pi$ it can be viewed as a 
presentation of the quantum double of some underlying infinite
dimensional Hopf algebra. The (TW) relations are then
characteristic for the intertwining operators between quantum double
modules \cite{FrResh,BeLeCl,KLP}. Particular cases are the Yangian double or 
the quantum double of $U_q(\hat{g})$ in which case the parameter $\beta$ 
can be related to the central extension via $\hbar c=i(2\pi -\beta)$
\cite{Iohara}. Here we do not make use of the co-algebra structure 
and always treat $\beta$ as a (real) numerical parameter entering the 
algebra via the set $I$ and the defining relations. The case of the 
``critical level'' with enlarged center in our conventions corresponds 
to $\beta =0$; it will be excluded throughout this paper without
further mentioning. A choice of conventions in particular amounts to fixing
a notation for $T^+$ and $T^-$, as they enter asymmetrically in 
$\cT_{\beta}(S)$. The flip isomorphism is given by
\be
\cT_{\beta}(S) \rra \cT_{4\pi -\beta}(S)\;,\;\;\;
T^{\pm}(\th)_a^b \rra T^{\mp}(\th+i\beta -i2\pi)_a^b\;.
\label{iso1}
\ee
Further (T1) has the usual consequences for the traces 
$t^{\pm}(\th) = T^{\pm}(\th)_a^a$ separately, i.e.
$$
[t^{\pm}(\th_1)\,,\,t^{\pm}(\th_2)]=0\;,
$$
but $t^+(\th_1)$ and $t^-(\th_1)$ will no longer commute for  
$\beta \neq 2\pi$.

There are two degree zero elements in $\cT_{\beta}(S)$ that do not 
appear in the defining relations, namely
\ba
&& C_{ab}(\th) := 
C_{mn}T^+(\th)_a^m\cdot T^+(\th+i\pi)_b^n \in \cA^{(1,0)}\nonum
&& C^{ab}(\th) := C^{mn}T^+(\th)_m^a\cdot 
T^+(\th-i\pi)_n^b \in \cA^{(1,0)}\;,
\label{c1}
\ea
and similar expressions in terms of $T^-$.. One easily verifies
that in general these are not central elements. However when the    
$S$-matrix is $2\pi i$-periodic -- and only then -- the 
following simplifications take place:
(i) The operators $C_{ab}(\th)$ and $C^{ab}(\th)$ reduce to 
$C_{ab}$ and $C^{ab}$, respectively. (ii) The operators 
$T^{\pm}(\th)_a^b$ are $2\pi i$-periodic. 
(iii) The relation (T1) can be assumed to hold also for 
Re$\,\th_{12} =0$. (iv) $D^-_{ab}(\th)$ as defined in (\ref{D1}) 
can be rewritten as 
\be
D^-_{ab}(\th)=C_{mn}\,T^+(\th)_a^m\cdot T^-(\th+i\pi)_b^n\;.
\label{D2}
\ee
To see point (i) note the relations
\ba
&& S_{ab}^{mn}(i\pi) C_{mn}(\th) = - \dim V\, C_{ab}\;,\nonum
&& S^{ab}_{mn}(i\pi) C^{mn}(\th) = - \dim V\, C^{ab}\;.
\label{c4}
\ea
Thus, provided $S_{ab}^{dc}(-i\pi)$ is well defined, -- which holds
in particular when the S-matrix is $2\pi i$-periodic -- one deduces
upon contraction 
\be
C_{ab}(\th) = C_{ab}\;,\sspace C^{ab}(\th) = C^{ab}\;.
\label{c5}
\ee
As a by-product one finds that the regularity of $S_{ab}^{dc}(-i\pi)$
is also a sufficient condition for the S-matrix to be $2\pi i$-periodic,
provided the relations (T1), (T2) hold (for Re$\,\th_{12} \neq 0$).
Indeed, if $S_{ab}^{dc}(-i\pi)$ is regular one concludes from 
(\ref{c5}) and (T2) that the operators $T^{\pm}(\th)_a^b$ are 
$2\pi i$-periodic. Since the relations (T1) with periodic $T^{\pm}$
operators only make sense if also the S-matrix is periodic,
one obtains the implication
\be
\mbox{(T1), (T2):}\sspace S_{ab}^{dc}(-i\pi)\;\;\mbox{is regular}\;\; 
\Longrightarrow \;\;
S_{ab}^{dc}(\th)\;\;\mbox{is $2\pi i$-periodic}\;.
\label{s7}
\ee
This in turn implies the dichotomy announced in (\ref{s5}).  
The point with (iii) is that assuming (T1) to hold also 
for $Re\,\th_{12} =0$ the $2\pi i$-periodicity of $T^{\pm}$ can be 
deduced from (T2). Thus, the former only makes sense if the latter
holds anyway, i.e. if the S-matrix is $2\pi i$-periodic. To see
the implication (iii) $\Rightarrow$ (ii) it suffices to show that
(iii) implies (\ref{c5}). Consider first $C_{ab}(\th) =
C_{mn}T^+(\th)_a^m \cdot T^+(\th +i\pi)_b^n$. Inserting 
$C_{mn} = - \dim V \,S_{mn}^{pq}(-i\pi) C_{pq}$ and 
using (T1) for $\th_1 = \th,\;\th_2 = \th +i\pi$, a further  
application of (T2) reduces the expression to $C_{ab}$. The second 
relation (\ref{c5}) is obtained similarly. Finally (iv) is a trivial
consequence of (iii).

Next we consider some implications of (M). The relation (M) allows 
one to compute the contraction product of $W$ generators at all 
relative rapidities $\th_{12} =i\pi +ip\beta$ and 
$\th_{12} =-i\pi +ip\beta,\;p\in\Z$ from (R). Using (M) and (TW) 
for one finds e.g.~for $p=1$ and generic $\beta$ 
\ba
&& W_a(\th +i\beta -i\pi)\cdot W_b(\th) = 
\lb\,D^+_{ab}(\th+i\beta -2i\pi)\;,\nonum
&& W_a(\th -i\beta +i\pi)\cdot W_b(\th) = 
\frac{\lb}{\dim V}\;D^-_{ab}(\th-i\pi)\;.
\label{res1}
\ea
For $\beta =2\pi$ a similar computation maps the two relations 
(R) onto each other. Observe also that for generic $\beta$ the 
contraction products (R) and (\ref{res1}) are consistent with an 
extension of the exchange relations (WW) to purely imaginary 
relative rapidities
\ba
&& W_a(\th + i\pi)\cdot W_b(\th) = - S_{ab}^{dc}(i\pi)\,   
W_c(\th)\cdot W_d(\th +i\pi)\;,\nonum
&& W_a(\th +i\beta -i\pi)\cdot W_b(\th) =
- S_{ab}^{dc}(i\beta-i\pi)\,W_c(\th)\cdot W_d(\th+i\beta -i\pi)\;.
\label{res4}
\ea
Anticipating the later interpretation of the WW contraction product as
a residue (within certain functionals), equations (\ref{res4}) are what 
one would expect if the exchange relations (WW) were valid also at 
purely imaginary relative rapidities.

Finally we note that the algebras $\cT_{\beta}(S),\;\cM_{\beta}(S)$ and 
$\cF_{\beta}(S)$ are equipped with a $*$-operation. In technical terms 
a $*$-operation is an antilinear anti-involution of some associative 
algebra. Here we denote such operations by $\sigma$ since $*$ 
is already used for complex conjugation and $\dagger$ would be 
cumbersome. The algebras  $\cT_{\beta}(S),\;\cM_{\beta}(S)$ and 
$\cF_{\beta}(S)$ admit an antilinear anti-involution $\sigma$ given by 
\be
\sigma T^{\pm}(\th)_a^b = T^{\mp}(\th^*+i\beta -i\pi)_a^b\;,
\sspace \sigma W_a(\th) =W_a(\th^*+i\pi)\;,\;\;\;\sigma^2 =id\;.
\label{invol}
\ee
Other $*$-products can be obtained from it by suitable basis 
transformations on $V$ and its dual that leave the S-matrix and the 
charge conjugation matrix invariant. Trivally (\ref{invol}) could also 
be modified by an overall shift $\th$ on the right hand side by a purely 
imaginary number. The particular choice (\ref{invol}) adheres to the 
crossing relations for the form factor equations. 

\newsubsection{T-invariant functionals and form factor sequences}

Consider now linear functionals $\omega_{\beta}:\cF_{\beta}(S)\ra \C$.
We call a linear functional $T$-{\em invariant} if it Lorentz invariant
$\omega_{\beta}(e^{i\lb K}\,X) =\omega_{\beta}(X\,e^{i\lb K}) =
\omega_{\beta}(X),\;\lb \in \R$ and satisfies
\be
\ob(T^-(\th)_a^b\,X) = \eta\,\delta_a^b\,\ob(X)\;,\sspace
\ob(X \, T^+(\th)_a^b) = \delta_a^b\,\ob(X)\;,
\label{state1}
\ee
for all elements $X \in \cF_{\beta}(S)$ with rapidities separated from 
$\th$. An element $X \in \cA^{(m,n)}$ depending on rapidities 
$\th_1,\ldots,\th_m$ is said to have rapidities separated from 
$\th$, if $\th_j -\th \not\in \Z I$, $1\leq j\leq m$. 
The functionals (\ref{state1}) are the form factor analogue of the 
thermal equilibrium states and will turn out to be invariant under the 
(counterpart of the) action of the modular operators. 
Important examples of such functionals are vector functionals 
(not traces!) 
\be 
\omega_{\beta}(X) = \Obra X \Oket\;,
\label{state2}
\ee
built from a {\em pair} of vectors $\Oket$ and $\Obra$
satisfying  
\be 
T^+(\th)_a^b \Oket = \delta_a^b \Oket \;,\sspace
\Obra T^-(\th)_a^b  = \eta \delta_a^b\Obra \;,
\label{state3}
\ee
where in general $\Obra \Omega_{\beta}\ket =0$. We shall later 
address the question to what extent these vector functionals are 
generic. Here observe that any $T$-invariant functional (\ref{state1}) 
is uniquely determined by its values on strings of W-generators, 
for which we introduce some extra notation
\be
f_{a_n\ldots a_1}(\th_n,\ldots,\th_1) := 
\ob(W_{a_n}(\th_n) \ldots W_{a_1}(\th_1))\;,
\label{state4}
\ee
where ${\rm Re}\,\th_{ij}\neq 0,\;i\neq j$.  Sometimes also the 
shorthand $f^{(n)}$ for the value of $\omega_{\beta}$ on a string of $n$ 
$W$-generators will be used. We can now partially restore topological 
concepts by calling a $T$-invariant functional {\em analytic} if: (i) 
the dependence of the values $f^{(n)}$ on the parameters 
$\th_1,\ldots,\th_n$ is locally analytic. (ii) overall shifts 
$\th_j \ra \th_j +i\beta/2$ leave the values $f^{(n)}$ invariant up 
to possibly a phase. (iii) $\ob(X\,W_a(\th_1)W_b(\th_2)\,Y)$ has simple 
poles at $\pm\th_{12} \in I$ whose residues are given by
$-i \ob(X\,W_a(\th_1)\cdot W_b(\th_2)\,Y),\;\pm\th_{12}\in I$ (possibly
contracted with $C^{ab}$), for elements $X,Y$ with rapidities separated 
from $\th_1,\th_2$. Further a $T$-invariant functional $\omega_{\beta}$ 
is called {\em hermitian} if 
\be
f_{a_1\ldots a_n}(\th_1^*+i\pi,\ldots,\th_n^*+i\pi) =
f_{a_n\ldots a_1}(\th_n,\ldots,\th_1)^*\;. 
\label{real2}
\ee
For generic elements  $X\in \cA^{(m,n)}$ this amounts to  
\be
\ob(\sigma(X)) = \eta^l\,\ob(X)^*\;,
\label{real1}
\ee
for some $X$-dependent integer $l$, which can be computed from 
(\ref{state1}) and (\ref{real2}). A $T$-invariant analytic and hermitian 
linear functional over $\cF_{\beta}(S)$ will be called a 
$T$-{\em invariant form}. Using the definition (\ref{state1}), and 
the relations of the algebra $\cF_{\beta}(S)$ one can write down a 
system of functional equations for the matrix elements (\ref{state4}) 
whose consistency is guaranteed by that of the underlying 
algebra $\cF_{\beta}(S)$.  
\medskip

\noindent {\bf Theorem 1:} {\em For any $T$-invariant 
form $\omega_{\beta}$ the sequences $(f^{(n)})_{n\geq 1}$ 
satisfy the coupled system} (I), (II) {\em of functional equations
described in appendix A.}
\medskip

The proof is a direct application of the defining relations \cite{MNalg}. 
The consistency of the resulting functional equations is ensured 
by the consistency of the underlying algebra. In particular 
any T-invariant functional over the modular algebra $\cM_{\beta}(S)$ 
produces solutions of the equivariance equations (I) for all $n \geq 1$. 
For $\beta = 2\pi$ the equations (I), (II) coincide with the 
form factor equations of a massive integrable QFT without bound 
states. For $\beta \neq 2\pi$ one gets a modified system of 
equations, and its solutions no longer describe the form factors of 
a standard QFT. The elements of a sequence $(f^{(n)})_{n\geq 1}$ 
will be called ``form factors'' for $\beta = 2\pi$ and ``deformed form
factors'' for $\beta \neq 2\pi$, or, when the distinction is
inessential, simply ``form factors''. The theorem then implies that each 
$T$-invariant form over $\cF_{\beta}(S)$ is uniquely determined by 
a sequence of form factors and vice versa, 
i.e. one has a 1--1 correspondence
\be
\omega_{\beta}\;\; \longleftrightarrow \;\;
(f^{(1)},f^{(2)},\ldots, f^{(n)}, \ldots )\;.
\label{1-1}
\ee
Comparing (\ref{real2}) with (\ref{I1}) one sees that hermitian linear 
forms correspond to form factors of hermitian operators. Theorem 1 also 
holds without this restriction, but for the moment we impose 
(\ref{real1}) for convenience.

To each $T$-invariant form $\ob$ a canonical quadratic form 
$(\;,\;)_{\omega}:\cF_{\beta}(S) \times \cF_{\beta}(S) \ra \R$ 
can be associated such that `off the diagonals', that is, 
whenever all rapidities of $X$ are separated from all of $Y$,
it obeys
\vspace{-2mm}
\be
(Y,X)_{\omega} = \omega_{\beta}(\sigma(Y)X)\;.
\label{quad1}
\ee
The explicit expression, valid also `on the diagonals', is relegated 
to appendix B. It is contravariant with respect to $\sigma$ and 
hermitian, i.e.
\be
(XY,Z)_{\omega} = (Y,\sigma(X)Z)_{\omega}\;,\sspace
(X,Y)_{\omega} = \eta^l\,(Y,X)_{\omega}^*\;,\;\;\; l\in \Z\;.
\label{quad2}
\ee
In general however the quadratic form (\ref{sesq1}) 
is not positive semi-definite, and cannot expected to be so 
on the grounds of (\ref{quad1}).

We call the $T$-invariant form $\ob$ {\em positive} if the associated 
quadratic form (\ref{sesq1}) is positive definite, i.e. $(X,X) >0$,
for all non-zero elements $X$ of the form (\ref{sesq0}). Heuristically one
expects $\ob$ to be positive precisely when the local operator whose 
form factors the sequence $f^{(n)}$ represents is positive.
Notice that we require non-degeneracy only for elements
of the form (\ref{sesq0}). On elements involving $T^{\pm}$ generators 
the quadratic form (\ref{sesq1}) is inevitably degenerate due to 
the T-invariance condition (\ref{state1}). 

Let us now address the issue under what conditions a T-invariant 
functional can be written as a vector functional (\ref{state2}).
Starting with a positive T-invariant form one clearly expects this 
to be the case. This is because one then is in the typical situation
where a GNS construction applies. Of course the GNS theorem does 
not apply literally (generically one is dealing with unbounded
operators and topological notions have only introduced indirectly).
Nevertheless the basic construction should still apply and 
yield a state space $\cH_{\omega}$ with cyclic vector $\Omega_{\omega}$
and a representation $\pi_{\omega}$ of $\cF_{\beta}(S)$ acting on it.
The vector $\Omega_{\omega}$ then is the mathematically well-defined  
version of the symbolic expression $(e^{2\pi K} \cO)^{1/2}$, where 
$\cO$ is a positive local operator in the QFT and $K$ is the 
generator of Lorentz boosts.
Of course positivity of $\omega$ here is essential because only then 
the state space $\cH_{\omega}$ will inherit the positivity of 
(\ref{sesq1}). In the representation $\pi_{\omega}$ other local
operators, not necessarily positive, should have a well-defined 
action on $\cH_{\omega}$. Of course the inner product of two vectors 
generated thereby can no longer directly set into correspondence to a 
form factor.
A way to maintain a correspondence to form factors would be to 
sacrifice positivity of the state space and to take 
(\ref{state3}) as the defining relation for a vector $\Oket$, whether
or not the associated functional (\ref{state2}) is positive.   
The state space associated with $\Oket$ then is 
\be
\Sigma = \bigoplus_{n\geq 0} \Sigma^{(n)}\;,\sspace
\Sigma^{(n)} = \bigoplus_{m\geq n}\cA^{(m,n)}\Oket\;,
\label{vec9}
\ee
and coincides with $\cH_{\omega}$ only when $X \ra \Obra X\Oket$ 
is positive. 

\pagebreak[2]
\newsection{Modular structures}

In this section algebraic counterparts $(j,\delta)$ of the modular 
structures featuring in the Tomita-Takesaki theory are derived.
As explained in the introduction they neatly unravel the 
`finite temperature structure' underlying the form factor approach. 
The modular structures $(j,\delta)$ will be affiliated with the 
quantum operator algebras $\cM_{\beta}(S)$ or $\cF_{\beta}(S)$
containing a generalized quantum double $\cT_{\beta}(S)$
and a ZF-type algebra $\cW(S)$ as distinguished subalgebras. Both are 
linked, in particular, by the crucial ``modular'' relation (M). 
Roughly speaking the quantum double $\cT_{\beta}(S)$ can be regarded
as `unphysical' in that its elements can be eliminated from
expectation values with T-invariant functionals. The $\cW(S)$ 
subalgebra in contrast is `physical' in that its expectation 
values define the form factors (or solutions of the equivariance 
equations (I) in the case of $\cM_{\beta}(S)$).
From the viewpoint of the cyclic form factor equation, one of the 
goals of the formalism here is to make sense out of thermal expectation
values over $\cW(S)$ with the unbounded `density operator' 
$\Delta = e^{\beta K}$, where $K$ is the generator of Lorentz boosts. 
In upshot, the modular relation (M) allows one to do precisely this.
Namely to implement Lorentz boosts with imaginary parameter on
$\cW(S)$ in terms of the `unphysical' $\cT_{\beta}(S)$ generators,
such that in particular the thermal state condition comes out
correctly.  

In more detail consider the following abelian automorphism group 
on $\cW(S)$
\be
\delta_n W_a(\th) = W_a(\th -i \beta n) \;,\;\;\;n \in \frac{1}{2}\Z \;,
\sspace\delta_n \delta_m = \delta_{n+m} = \delta_m \delta_n\;.  
\label{MI1}
\ee
We set $\delta := \delta_1$. Formally $\delta_n = D_{-in\beta}$, if 
$D_{\lb}(W_a)(\th) = e^{i\lb K}W_a(\th) e^{-i\lb K}$, $\lb \in \R$ are 
the Lorentz boosts. The main results are:
\begin{itemize}
\item[(a)] $\delta_n W_a(\th),\;n\in\frac{1}{2}\Z$, can be 
implemented in terms of the $\cT_{\beta}(S)$ generators.
\item[(b)] The algebras $\cM_{\beta}(S)$ and $\cF_{\beta}(S)$ 
both `decompose' into a subalgebra $\cN$ and its commutant $\cN'$, 
which are related by an involution $j$ (Theorem 2A and 2B).
$\cN$ contains $\cW(S)$ as a subalgebra.  
\item[(c)] The operators $(j,\delta)$ have all algebraic
features of ``modular structures'' in the context of the 
Tomita-Takesaki theory (Theorem 2C). The counterpart of the 
``KMS property'' is
\be
\ob(Y\delta_1X) = \eta^{l} \ob(XY)\;,
\sspace X,Y \in \cN\,,\;l\in \Z\;,
\label{MI2}
\ee 
and generalizes the cyclic form factor equation.
\end{itemize} 
Let us add a few remarks. The `doubling of the degrees of freedom' in 
(b) is characteristic of finite temperature equilibrium 
dynamics. It is intimately linked to the modular relation (M) 
and the cyclic equation satisfied by the functions (\ref{state4}).
For clarities sake let us state that this holds irrespective of the
contractions (R) and hence in particular is true for the modular
algebra $\cM_{\beta}(S)$. Probably we should also repeat that one is 
not dealing with von 
Neumann algebras. The ``modular structures'' described  here can 
therefore not directly be subsumed into the framework of the 
Tomita-Takesaki theory (although this may turn out to be the case 
in an appropriate reformulation). Nevertheless the ``modular structures'' 
described here have all the {\em algebraic} features typical for 
modular structures in the context of von Neumann algebras.
Keeping in mind this disclaimer our borrowing of the terms
``modular conjugation'' and ``modular operator'' should not give 
rise to confusion.

Finally it may be worthwhile to point out the similarity to 
the ``corner transfer matrix formalism'' used in the context of 
integrable lattice models. A remarkable feature of the modular
operator $\Delta$ in the Tomita-Takesaki theory 
is that it plays a double role. On the one hand it implements the 
KMS condition via $(\Omega, Y \Delta X\Omega) = (\Omega, XY\Omega),\;
X,Y \in \cM$. On the other hand it can be used to define a unitary 
automorphism group, the modular `time evolution', which leaves the 
algebra $\cM$ and its commutant $\cM'$ separately invariant. Defining 
$K \sim \ln \Delta$ to be the ``modular hamiltonian'' one has 
$D_{i\lb}(X) = e^{i\lb K}X e^{-i\lb K},\;\lb\in \R$, for this automorphism 
group. In the context of form factors, the von Neumann algebra  
$\cM$ is replaced with the quantum operator algebra $\cN$ and 
the group $X \ra \Delta^n X \Delta^{-n}$ ($X$ analytic w.r.t. 
${\rm Ad}\Delta$) corresponds to 
$\delta_n$ in (\ref{MI1}). According to (a) the latter  
can be implemented in terms of the generalized quantum double
$\cT_{\beta}(S)$. The full analogue of the Tomita Takesaki theorem 
would state that $\delta_{i\lb/\beta}$, $\lb \in \R$, defines
an automorphism group for both $\cN$ and its commutant $\cN'$.
Let us suppose for the moment that such an analogue has been 
obtained. Then there are {\em two} ways of implementing 
Lorentz boosts: The `ordinary one' where $K$ is just a given 
generator of the (`kinematical') Poincar\'{e} group;
and a second one via $\delta_{it},\; t\in \R$, that is in terms of 
the (`dynamical') form factor algebra $\cF_{\beta}(S)$. A similar 
phenomenon has been discovered around 1979 by Baxter \cite{Baxter} in 
the context of lattice models, where the ``corner transfer matrix''
\cite{Baxter,Davies,Thacker} plays the kinematical/dynamical double 
role of $\delta$. Combined with the powerful techniques now available 
in the representation theory of infinite dimensional quantum algebras 
this lead to considerable progress in this area; see \cite{JimMiwa1} 
for an overview. On the QFT level the implementation of Lorentz boosts 
with imaginary parameter in terms of $\cT_{\beta}(S)$ may be viewed as 
as a counterpart of the corner transfer matrix formalism.


\newsubsection{Subalgebra $\cN$}

As mentioned before the algebras $\cM_{\beta}(S)$ and $\cF_{\beta}(S)$
differ only insofar as in the latter the contraction product of two 
W-generators is declared by (R). For the subalgebras 
$\cN \subset \cM_{\beta}(S)$ and $\cN \subset \cF_{\beta}(S)$ 
described below we therefore use the same symbol, keeping in mind that 
the latter differs only by one extra relation. Of course $\cN$ is a
shorthand for $\cN_{\beta}(S)$. 

We denote by $\cN\subset \cM_{\beta}(S)$ the subalgebra generated
by $W_a(\th),\;D^+_{ab}(\th)$ as defined in and (\ref{D1}). 
The fact that these elements form a subalgebra is manifest from 
the following relations: 
\vspace{4mm}
\eql{N1}
\bas \jot5mm
&& \sspace S_{ab}^{mn}(\th_{21})S_{mc}^{lk}(\th_{21}+i\beta -i\pi)\;
D^+_{nk}(\th_1)\,D^+_{ld}(\th_2) \nonum
&&\bspace\sspace = S_{cd}^{mn}(\th_{12})
S_{bn}^{kl}(\th_{12} +i\beta -i\pi)\;
D^+_{al}(\th_2)\,D^+_{km}(\th_1)\;.
\eas
\vspace{-2mm}
\eql{N2}
\bas \jot5mm
C^{mn}D^{\pm}_{nk}(\th) D^{\mp}_{lm}(\th+ i\pi) C^{kl} = \dim V\;,
\eas 
where $D^-_{ab}(\th)$ is the shorthand (\ref{D1}). Further (WW) and 
\vspace{4mm}
\eql{NW}
\bas \jot5mm
\;\;\;S_{ab}^{mn}(\th_{21}+i\beta -i2\pi)\,D^+_{cn}(\th_1)\,W_m(\th_2)
\is S_{ca}^{mn}(\th_{12} +i\pi) \,W_n(\th_2)\,D^+_{mb}(\th_1)\;,
\eas
where (N1) and (NW) both hold for all relative rapidities. The
analogue of (M) is 
\vspace{2mm}
\eql{NM}
\bas \jot5mm
&& W_a(\th) = -\frac{1}{\dim V}\,C^{mn} D^-_{am}(\th +i\beta -i2\pi)
W_n(\th +i\beta) \;,\nonum
&& W_a(\th +i\beta) = -\frac{1}{\dim V}\,C^{mn} W_m(\th) 
D^-_{na}(\th +i\beta - i\pi) \;.
\eas
For the action of $\sigma$ one finds 
\be
\sigma D^+_{ab}(\th) = D^+_{ba}(\th^* +i\beta -i2\pi)\;,
\ee
so that $\cN$ is also a $*$-subalgebra of $\cM_{\beta}(S)$. 
The subalgebra $\cN \subset \cF_{\beta}(S)$ is the subalgebra 
generated by $W_a(\th),\;D^+_{ab}(\th)$ subject to the above relations
and (R).

Before proceeding let us remark that the form factor equations 
(I), (II) could not have been formulated for functionals over the 
subalgebra $\cN$ only. The appropriate invariance condition on such
functionals still  would have to guarantee that they are fully 
determined by their values on strings of W-generators. 
The T-invariance condition (\ref{state1}) used in section 2 is too weak. 
(For example a $T^-$- generator arising through $\ob(X\,D^+_{ab}(\th))= 
C_{bb'}\ob(X\,T^-(\th +i\pi)_a^{b'})$ is not an element of $\cN$ and 
cannot be pushed to the left by using the relations of $\cN$ only.) 
On the other hand a stronger invariance condition adapted to (NM)
would no longer reproduce the deformed KZE. We will see 
below that both $\cN$ and its commutant in $\cF_{\beta}(S)$ or
$\cM_{\beta}(S)$ are needed.


\newsubsection{Modular conjugation and modular operator}

Let $\cbF_{\beta}(S)$ denote the algebra $\cF_{\beta}(S)$ 
with the following replacement of `structure constants' 
\be
S_{ab}^{dc}(\th) \rra [S_{ab}^{dc}(\th)]^*\;,\sspace
\lb \rra -\lb\;.
\label{j1}
\ee
Further let $\cbN\subset \cbF_{\beta}(S)$ denote the counterpart of 
the subalgebra $\cN\subset \cF_{\beta}(S)$.  
Similarly let $\cbM_{\beta}(S) = \cM_{\beta}(S^*)$ be the 
modular algebra with the complex conjugate S-matrix and 
$\cbN \subset \cbM_{\beta}(S)$ the corresponding subalgebra. 
Of course one expects that the original algebras and their 
`barred' counterparts are basically the same, i.e.~isomorphic.
Indeed, a trivial isomorphism is the one that acts as the identity
on operators and as complex conjugation on complex numbers. 
A much more interesting isomorphism is described in Theorem 2A;
what makes it interesting is the content of parts B and C of 
Theorem 2. 

To describe this isomorphism let us prepare extra symbols 
$\Wbar_a(\th),\;\Tbar^{\pm}(\th)_a^b$ for the 
generators of $\cM_{\beta}(S)$ and $\cbF_{\beta}(S)$. Similarly we use 
$\Wbar_a(\th)$ and $\Dbar^+_{ab}(\th)$ generators of the respective 
$\cbN$ subalgebras. Define an anti-linear operator $j$ acting on 
$\cM_{\beta}(S)$ and $\cF_{\beta}(S)$ by 
\begin{subeqnarray} 
jT^+(\th)_a^b &\,=\,&  C_{aa'}C^{bb'} \,
T^+(\th^* + \frac{3}{2} i\beta - i2\pi)_{b'}^{a'}\;,\nonum
jT^-(\th)_a^b  &\,=\,& C_{aa'}C^{bb'} \,
T^-(\th^* + \frac{1}{2} i\beta )_{b'}^{a'}\;,\\
jW_a(\th)  &\,=\,&  C_{aa'} C^{mn} 
W_m(\th^* -\frac{1}{2}i\beta +i\pi)\cdot T^+(\th^*
+\frac{1}{2}i\beta)_n^{a'}\nonum
  &\,=\,& C_{aa'}C^{mn} T^-(\th^* +\frac{1}{2} i\beta)_n^{a'}\cdot  
W_m(\th^* + \frac{1}{2}i\beta + i\pi)\;,
\label{j2}
\end{subeqnarray}
and by
\be
j(XY) =j(X)j(Y)\;,\;\;\;j(z X) = z^* j(X)\;,\;\;\;z\in \C
\label{j3}
\ee
on products of generators. 
\newpage

\noindent{\bf Theorem 2A:}
{\em 
\begin{itemize}
\item[(1)] $j:\cM_{\beta}(S) \rra \cbM_{\beta}(S)$ and 
$j:\cF_{\beta}(S) \rra \cbF_{\beta}(S)$ defined by 
\be
\Wbar_a(\th) = jW_a(\th)\;,\sspace 
\Tbar^{\pm}(\th)_a^b = j T^{\pm}(\th)_a^b\;,
\label{j4}
\ee
are anti-linear isomorphisms of $*$-algebras and involutions,
i.e. $j^2 = id$.
\item[(2)] $j:\cN \rra  \cbN$ is an anti-linear 
isomorphism of $*$-subalgebras for both $\cN \subset\cM_{\beta}(S)$
and $\cN \subset \cF_{\beta}(S)$.
\end{itemize}
}
{\em Proof:} {\em (1)} One has to check that $\Wbar_a(\th)$ 
and $\Tbar^{\pm}(\th)_a^b$ as defined through (\ref{j2}), (\ref{j4}) 
satisfy all the relations of $\cbM_{\beta}(S)$ and $\cbF_{\beta}(S)$. 
For $\cbM_{\beta}(S)$ this can be verified by direct
although tedious computation. It remains to check the contraction
products for the W-generators. Using the definitions (\ref{j2}) of 
and (TW) one computes for generic $\beta$
\ba 
&& jW_a(\th +i\pi)\cdot jW_b(\th) = \lb\,C_{ab}\;,\nonum
&& C^{ab}jW_a(\th -i\pi)\cdot jW_b(\th) = \lb\;.
\label{j6}
\ea
These are the contraction products for $\cbF_{\beta}(S)$, 
as asserted. Consistency requires that also the contractions (\ref{res1})
come out correctly. Indeed, using again the definitions 
(\ref{j2}) of $jW_a(\th)$ one can verify
\ba
&& jW_a(\th +i\beta -i\pi)\cdot jW_b(\th)= 
- \lb\,jD^+_{ab}(\th+i\beta -2i\pi)\;,\nonum
&& jW_a(\th +i\pi-i\beta )\cdot jW_b(\th)= 
\frac{-\lb}{\dim V}\;jD^-_{ab}(\th-i\pi)\;.
\label{j8}
\ea
On the other hand the same contractions can be computed directly in terms 
of the barred generators starting from (\ref{j6}) and the relation (M)
in $\cM_{\beta}(S)$. The result is just (\ref{j8}), consistent with 
the identifications (\ref{j4}). For $\beta =2\pi$ one proceeds as 
follows. Using both versions of (\ref{j2}b), applying (R) and then
simplifying one obtains
\ba
jW_a(\th +i\pi)\cdot jW_b(\th) \is
C_{aa'}C_{bb'} C^{mn} T^-(\th^*)_n^{a'} \cdot W_m(\th^* + i\pi)\cdot
W_k(\th^*)\cdot T^+(\th^* + i\pi)_l^{b'} 
\nonum
\is \lb[C_{ab} - jD_{ab}^+(\th)]\;,
\label{j15}
\ea
as required. From here $C^{ab}jW_a(\th -i\pi)\cdot jW_b(\th) = \lb - \lb 
C^{ab}D^-_{ab}(\th -i\pi)/\dim V$ is verified using $j$(M) and $j$(TW). 
One can also check that this evaluation procedure is compatible with the 
the condition $0 \leq {\rm Im}\,\th \leq \pi$ (actually enforcing
$\th \in \R$ in (\ref{j15})) and that it is the 
only one. Finally the property $j^2= id$ can be verified. 
The statement {\em (2)} is a direct consequence of {\em (1)} and 
equations (\ref{j6}), (\ref{j8}). $\Box$ 

The most remarkable property of $j$ is that the original 
generators commute with all their $j$-transformed counterparts.
We formulate this as a Lemma because we shall later encounter a 
stronger version thereof.
\medskip

{\bf Lemma:} The generators of $\cN$ and $j(\cN)= \cbN$ mutually
commute, for both $\cN \subset\cM_{\beta}(S)$
and $\cN \subset \cF_{\beta}(S)$. Explicitly, for $Re\,\th_{12}\neq 0$
\begin{subeqnarray}
&& j W_a(\th_1)\,W_b(\th_2) = W_b(\th_2)\,j W_a(\th_1)\;,\\
&& j W_a(\th_1) \,D_{bc}^{\pm}(\th_2) = 
D_{bc}^{\pm}(\th_2)\,j W_a(\th_1)\;,\\
&& j D^{\pm}_{ab}(\th_1) \,D_{cd}^{\pm}(\th_2) =
D_{cd}^{\pm}(\th_2)\,j D^{\pm}_{ab}(\th_1)\;,
\label{j5}
\end{subeqnarray}
where in (c) all combinations of the $\pm$ options are allowed.
\medskip

The proof is by direct computation. To proceed let 
$\delta_n,\,n\in \Z$, denote the following discrete automorphism 
group of $\cM_{\beta}(S)$ or $\cF_{\beta}(S)$
\ba
&& \delta_n T^{\pm}(\th)_a^b = T^{\pm}(\th - i \beta n)_a^b\;,\sspace
\delta_n W_a(\th) = W_a(\th -i \beta n) \;,\;\;\;n \in \frac{1}{2}\Z \;,
\nonum
&& \bspace\sspace\;\;\;\;
\delta_n \delta_m = \delta_{n+m} = \delta_m \delta_n\;. 
\label{delta1}
\ea
Formally $\delta_n = D_{- i n \beta}$ if $D_{\lb}(X(\th)) = e^{i\lb K}
X(\th) e^{-i\lb K} = X(\th + \lb)$, $\lb \in \R$ are the Lorentz boosts. 
As emphasized in the introduction $D_{- i n \beta}$ is however meaningless  
when defined in terms of the generator $K$ of the Lorentz boosts. 
On the `unphysical' $\cT_{\beta}(S)$ subalgebra we are free to 
stipulate that $T^{\pm}(\th)_a^b$ is well defined for generic complex 
$\th$. Since they can be eliminated from the `physical' expectation
values $\ob(X)$ no meaning has to be given to $T^{\pm}(\th - in \beta)_a^b$
in terms of Lorentz boosts with an imaginary parameter. Of course the 
opposite is true for the subalgebra $\cW(S)$ generated by the W-operators.
In the present framework the action of $D_{- i n \beta}$ on $\cW(S)$ 
can however be implemented in terms of the $\cT_{\beta}(S)$ generators, 
exploiting once more the crucial relation (M).
This holds for both the modular algebra $\cM_{\beta}(S)$ and the 
form factor algebra $\cF_{\beta}(S)$. Explicitly 
\smallskip
\ba
&& \delta_{-1} W_a(\th) = C_{mn}T^-(\th +i\beta)_a^m \cdot W_k(\th)
                   \cdot \sigma T^-(\th^*)_l^n C^{lk}\;,\nonum
&& \delta_{1} W_a(\th) = C^{kl}\sigma T^+(\th^*+i\beta)_l^n \cdot 
                W_k(\th)\cdot T^+(\th- i 2\pi)_a^m C_{mn}\;,
\label{delta2}
\ea
\smallskip
from (\ref{w1}) and (\ref{invol}). By iteration $\delta_n,\,n\in \Z$, 
yields a discrete automorphism group of the $\cW(S)$ subalgebras of both  
$\cF_{\beta}(S)$ and $\cM_{\beta}(S)$.

So far $\delta_n$ has only been defined for integer $n$. 
Also the square root of $\delta_1$, say, can be 
implemented in terms of the $T^{\pm}$-generators and $j$.  
It is given by the following alternative expressions
\ba
\delta_{1/2} W_a(\th) \is \left\{ \begin{array}{l}  
j W_n(\th^* + i\pi) \cdot T^+(\th + \frac{1}{2}i\beta -i2\pi)_a^n \\
T^-(\th - \frac{1}{2} i\beta)_a^n \cdot j W_n(\th^* + i\beta +i\pi)\;.
\end{array}
\right. 
\label{delta3}
\ea 
Using the relation (M) both expressions can be seen to give 
$\delta_{1/2} W_a(\th) = W_a(\th -\frac{1}{2}i\beta)$ as required 
by consistency. In this sense (\ref{delta3}) it is not an independent 
automorphism. However, given the W-generators in a suitable strip
of the complex $\th$ plane (e.g.~${\rm Im}\,\th \in [0,\pi]$ for the 
first expression) as well as $j$, equation (\ref{delta3}) can be 
used to extend the domain of definition by half-integer multiples of 
$i\beta$, as it should.

The relation of (\ref{delta3}) to the square root of $\delta$ 
can also be seen as follows. There exists a linear (not antilinear) 
anti-automorphism $\sbar$ of $\cF_{\beta}(S)$ and $\cM_{\beta}(S)$ 
given by \cite{MNalg} 
\smallskip
\ba
&&\sbar T^+(\th)_a^b =C_{aa'}C^{bb'}\,
T^-(\th +i\pi)_{b'}^{a'}\;,\nonum
&&\sbar T^-(\th)_a^b =C_{aa'}C^{bb'}\,
T^+(\th +i\beta -i\pi)_{b'}^{a'}\;,\nonum
&& \sbar W_a(\th) =  W_m(\th)\cdot \sbar T^-(\th)_a^m=
   \sbar T^+(\th+i\beta -2\pi i)_a^m\cdot W_m(\th+i\beta)\;.
\label{delta4}
\ea
The proof is by direct computation, i.e.~stipulating that $\sbar(XY) = 
\sbar(Y)\sbar(X)$ holds, one checks that the transformed generators 
again satisfy all relations of $\cF_{\beta}(S)$ or $\cM_{\beta}(S)$.  
The square of $\sbar$ is 
\be
\sbar^2 T^{\pm}(\th)_a^b = T^{\pm}(\th +i\beta)_a^b\;,\sspace 
\sbar^2W_a(\th) = W_a(\th+i\beta)\;,
\label{delta5}
\ee
suggesting already a relation to the square root of the 
$\th \ra \th \pm i\beta$ automorphism. To unravel it observe that
also $\sbar$ has the commutant property described in the Lemma \cite{MNalg}: 
The generators of $\cN$ and $\sbar(\cN)$ mutually commute, for both 
$\cN \subset\cM_{\beta}(S)$ and $\cN \subset \cF_{\beta}(S)$. 
This ensures consistency of the following relation between
$\sbar$, $j$ and $\sigma$: $\sbar \circ \sigma = j \circ \delta_{1/2}$ 
$(*)$, where $\delta_{1/2}$ is presently simply a shorthand for the 
shift operation $\th \ra \th - i\beta/2$. The point here is that 
$\sbar$, $j$ and $\sigma$ are already known to be genuine 
(anti-)automorphisms of $\cM_{\beta}(S)$ and $\cF_{\beta}(S)$, so 
that solving $(*)$ for $\delta_{1/2}$ yields the searched for automorphism  
\be
\delta_{1/2} =  j\circ \sbar \circ \sigma \;.
\label{delta6} 
\ee
Evaluating it on $W_a(\th)$ using (\ref{invol}), (\ref{delta4})
and (\ref{j2}) gives (3.14). For the $\cT_{\beta}(S)$
generators one obtains $\delta_{1/2} T^{\pm}(\th)_a^b = 
T^{\pm}(\th - i \beta/2)_a^b$ as it should. Consistency requires 
that $(\delta_{1/2})^2 = \delta_1$, which follows from 
$\sigma\circ \sbar \circ \sigma = \sbar \circ \delta_1 =\sbar^{-1} =
j \circ \sbar \circ j$.  
\smallskip

In summary a discrete abelian automorphism group $\delta_n,\;
n \in \frac{1}{2}\Z$, on $\cM_{\beta}(S)$ and $\cF_{\beta}(S)$ has
been defined that implements `Lorentz boosts with imaginary parameter'
$D_{-in\beta}$ on the `physical' $\cW(S)$ subalgebra consistently 
in terms of the `unphysical' $\cT_{\beta}(S)$ generators.
The T-invariance condition (\ref{state1}) can now be re-interpreted 
as the invariance under this discrete automorphism group 
\be
\ob(\delta_n(X)) = \eta^l\,\ob(X)\;,\sspace n \in 
\frac{1}{2}\Z\;,
\label{delta7}
\ee
for some $l\in \Z$. For integer $n$ this follows algebraically 
from the defining relations, for half-integer $n$ it is a consequence 
of the postulated analyticity properties of (\ref{state4}).  
\smallskip

Next we turn to the interplay between the automorphism group 
$\delta_n$ and the involution $j$.
\pagebreak[2]

{\bf Theorem 2B:} {\em Let $\ob$ be a T-invariant form on either 
$\cF_{\beta}(S)$ or $\cM_{\beta}(S)$.  
\begin{itemize}
\item[(1)] $j \circ \sigma = \sigma \circ j$ and    
$j \circ \delta_n = \delta_{-n} \circ j$ and $\sigma \circ \delta_n =
\delta_{-n} \circ \sigma$. 
\item[(2)] $j$ is anti-unitary (up to possibly a phase) with respect 
to the quadratic form (\ref{sesq1}) induced by $\ob$ and $\sigma$, i.e. 
$ (j(X),j(Y)) = \eta^l(Y,X)\;,\;l\in \Z\;.$
\item[(3)] Let $X \in \cN$ and $Y$ an element of $\cF_{\beta}(S)$ or
$\cM_{\beta}(S)$ with rapidities separated from that of $X$. Then   
$$
\ob(Y\,(j \circ \delta_{1/2})(X)) = \ob(Y\,\sigma(X))\;,\;\;
\ob((j \circ \delta_{-1/2})(X)\,Y) = \eta^l\ob(\sigma(X)\,Y)\;,\;
\;\;l\in \Z\;. 
$$
\end{itemize}
}
\medskip

{\em Proof:} (1) and (3) are verified by direct computation.
Let us illustrate (3) for the case of the $\cW(S)$ subalgebra of $\cN$.
To simplify the notation take $\ob$ to be a vector functional built
from a T-invariant vector $\Oket$. Using the Lemma one finds  
\be
j[W_{a_n}(\th_n) \ldots W_{a_1}(\th_1)] \Oket =
W_{a_1}(\th_1^* + i\pi -\frac{1}{2} i\beta ) \ldots 
W_{a_n}(\th_n^* + i\pi -\frac{1}{2} i\beta )\Oket\;,
\label{delta8}
\ee 
so that $j(X)\Oket = (\delta_{1/2}\circ \sigma)(X)\Oket$ for 
$X \in \cW(S)$. The same can be checked for the generator 
$D^+_{ab}(\th)$ of $\cN$ and generic mixed products.   
Since only the `ket' T-invariance condition was used 
an equivalent way to present the result is 
$\ob(Y\,(j \circ \delta_{1/2})(X)) = \ob(Y\,\sigma(X))$, as asserted.
For (2) it suffices to consider the case where all rapidities of 
$X$ are separated from all of $Y$, so that $(Y,X)$ in (\ref{sesq1}) 
reduces to $\ob(\sigma(Y)X)$. The anti-unitarity of $j$ then amounts
to $\ob(j(Z)) = \eta^l \ob(\sigma(Z))$ with $Z= \sigma(X)Y$, which 
follows from part (3) and (\ref{delta7}). $\Box$

The properties of $j$ and $\delta_n$ described in Theorem 2B clearly 
parallel that of the modular structures $(J,\Delta)$ in the 
Tomita-Takesaki theory of von Neumann algebras (e.g.~as outlined in the 
introduction). The identifications are $j(X) \ra JXJ$ and  $\delta_n(X) 
\ra \Delta^n X \Delta^{-n}$ ($X$ analytic w.r.t. ${\rm Ad} \Delta$), 
where $\Delta$ defines the generator of the 
Lorentz boosts by $\Delta = e^{\beta K}$. Motivated 
by this analogy we shall refer to $j$ as the {\em modular conjugation} 
and to $\delta$ as the {\em modular operator} of the form factor
algebra $\cF_{\beta}(S)$ or the modular algebra $\cM_{\beta}(S)$.
The little computation (\ref{I7}) establishing the ``KMS property'' 
of the modular operator $\Delta$ can directly be taken over:
Let $X,Y$ elements of $\cN$ with rapidities separated from each other,
so that  $(Y,X)$ in (\ref{sesq1}) reduces to $\ob(\sigma(Y)X)$. 
Let $l_1,l_2$ be suitable integers. Then
\ba
\ob(Y \delta_1 X) \is \eta^{l_1} \,
\ob\left(\delta_{-1/2}(Y) \delta_{1/2}(X)\right) =
\eta^{l_1} \left( \sigma\delta_{-1/2}Y, j(\sigma X)\right)
\nonum \is 
\eta^{l_1+l_2}\,\left(\sigma X,j \sigma(\delta_{-1/2} Y)\right) 
= \eta^{l_1+l_2}\,\ob(XY)\;.
\label{delta9}
\ea
Thus 
\be
\ob(Y\delta_1X) = \eta^l \,\ob((\delta_{-1}Y) X) = \eta^{l'} \ob(XY)\;,
\sspace X,Y \in \cN\;,
\label{delta10}
\ee 
with $l,l' \in \Z$. In particular for $Y= W_{a_n}(\th_n),\;
X= W_{a_{n-1}}(\th_{n-1})\ldots W_{a_1}(\th_1)$ one recovers the cyclic 
form factor equation. Of course this is also follows directly from (M) and 
(\ref{state1}) (see Theorem 1 and \cite{MNalg}), but it is gratifying to 
see it reappear from the underlying `finite temperature' automorphism 
structure, without having to appeal to the deformed KZE equation.

In order to complete the analogy with the Tomita-Takesaki theory,
the image $j(\cN)$ of the subalgebra $\cN$ should coincide with the 
commutant of $\cN$ in $\cF_{\beta}(S)$ or $\cM_{\beta}(S)$.
A little technical problem here is that 
$j(\cN)$ lives in $\cbF_{\beta}(S)$ or $\cbM_{\beta}(S)$, 
while $\cN$ is a subalgebra of the `unbarred' algebras.  
This can easily be rectified by concatenating $j$ with the 
trivial anti-linear automorphism $\iota$ implementing the flip
(\ref{j1}) and acting like the identity on algebra elements. 
Denoting by $\jmath = \iota \circ j$ this concatenation,
$\jmath(\cN)$ is a subalgebra of  $\cF_{\beta}(S)$ or $\cM_{\beta}(S)$
of which one can ask how it relates to the original subalgebra $\cN$.
Next one has to specify what one means by the commutant of 
$\cN$ in this context. Naturally one will require that $\cN'$
is again a quantum operator algebra generated from degree 1 elements
$X(\th)$ by the two multiplication operations. 
Further it should be the largest quantum operator subalgebra 
of $\cF_{\beta}(S)$ or $\cM_{\beta}(S)$ commuting with $\cN$.
To simplify matters we also assume here that the S-matrix is
$2\pi i$-periodic. Then both $W_a(\th \pm i\pi)\cdot W_b(\th)$ products
enter symmetrically and the identity (\ref{c5}) simplifies the 
structure of the commutant. With these specifications the following 
result holds. 
\medskip

{\bf Theorem 2C:}
{\em
For a $2\pi i$-periodic S-matrix let 
$\cN'$ denote the commutant of $\cN$ in $\cF_{\beta}(S)$.
Then $\cN'$ is a $*$-subalgebra of $\cF_{\beta}(S)$ isomorphic 
to $\jmath(\cN)$. The same holds for $\cN \subset \cM_{\beta}(S)$. 
}

{\em Proof:} Lemma 3 implies that $\jmath W_a(\th)$ and  
$\jmath D^{\pm}_{ab}(\th)$ and hence all 
finite products thereof are elements of the commutant of $\cN$, 
symbolically $\jmath(\cN) \subset \cN'$. Theorem 2A implies that they 
generate a quantum operator algebra isomorphic to $\iota(\cbN)$. On the 
other hand the definitions (\ref{state3}), (\ref{j2}) entail that $\cN$ 
and $\jmath(\cN)$ generate the same state space (\ref{vec9}). Since 
$\jmath(\cN) \subset \cN'$ the same holds a-fortiori for $\cN'$. Thus 
\be
\Sigma = \cN \Oket = (\jmath\cN)\Oket = \cN'\Oket\;.
\label{j9}
\ee 
From here on one can adopt a standard argument to conclude that 
\be 
\jmath(\cN) = \cN'\;\;\;(\mbox{as sets})\;,
\label{j10}
\ee
i.e.~all elements of $\cN'$ arise as images under $\jmath$:
Set $s = j \circ \delta_{1/2}$, which is an algebra-homomorphism and can 
be checked to satisfy $s^2 = id$ and $s(X) \Oket = \sigma(X)\Oket$, 
for all $X\in \cN$. By (\ref{j9}) one knows that for any $X\in \cN$ 
there exists an $X' \in \cN'$ such that $\sigma(X)\Oket = X' \Oket$. 
For any $Y\in \cN$ one then has 
\be
s(X) Y\Oket = s(X\sigma(Y))\Oket = Y\sigma(X)\Oket =
Y X'\Oket = X' Y\Oket\;.
\label{j11}
\ee
Hence $s\cN \subset \cN'$ and by a symmetric argument 
$s\cN' \subset \cN$. It follows that $s\cN = \cN'$ and 
thus also that $\jmath(\cN)$ and $\cN'$ coincide as sets. Since 
$\jmath(\cN) = \iota(\cbN)$ is already known to be a
$*$-subalgebra of $\cF_{\beta}(S)$ the same holds for $\cN'$. 
$\Box$  

At the expense of somewhat complicating the algebraic structure
this result can also be extended to non-periodic S-matrices.  
Let us emphasize that the result does not apply to elements
of `infinite degree', which are beyond the scope of the ``quantum 
operator algebra'' concept. 
Theorem 2C can be regarded as part of the form factor counterpart
of the Tomita-Takesaki theorem. The second part would consist 
in showing that starting from the definitions (\ref{delta2}), 
(\ref{delta3}) also $\delta_{i\lb/\beta}$, $\lb\in \R$ can be defined and
provides an automorphism group of both $\cN$ and $\cN'$ that 
coincides with the original Lorentz boosts $D_{\lb}$.    
Since $\cT_{\beta}(S)$ via the S-matrix carries dynamical information,
this would reveal part of its `kinematical -- dynamical' double 
role alluded to in the introduction to this section. 

\newsubsection{Further properties}

The usefulness of the automorphism 
\be
s= j\circ \delta_{1/2}=
\sbar \circ \sigma
\label{delta11}
\ee
encountered twice in the previous section is not 
accidental.%
\footnote{We apologize for not avoiding the clash of several standard 
notations here: $s$ must not be confused with the ``antipode'' in the 
context of Hopf algebras. The Tomita operator $\check{S}$ 
of course has nothing to do with the S-matrix. Both $s$ and $\check{S}$ 
in the sense of (\ref{delta11}), (\ref{delta12}) appear only within 
the next two paragraphs. In section 3.3 we write $\cN_{\beta}(S)$
for the quantum operator algebras of section 3.1 and 3.2.}
It is the counterpart of the Tomita operator $\check{S}$ forming 
the starting point of the Tomita-Takesaki theory. To explain this, let 
us briefly recap how the modular operators $(J,\Delta)$ are 
usually constructed in the context of von Neumann algebras:
Let $\cN$ be a von Neumann algebra in ``standard form''. This means 
that $\cN$ acts on a Hilbert space $\cH$ possessing a cyclic and 
separating vector $\Omega$ and that both $D :=\cN \Omega$ 
and $D' := \cN' \Omega$ are dense in $\cH$. Here as usual $\cN'$
denotes the commutant of $\cN$ in $\cB(\cH)$, the bounded operators 
on $\cH$. Define operators $\check{S}$ and $\check{F}$ by 
\ba
\check{S}: D\ra D\;,&\;\;\;& \check{S} X \Omega = X^* \Omega 
\sspace \;X\in \cN\;,
\nonum
\check{F}: D'\ra D'\;,&\;\;\;& \check{F} X' \Omega = (X')^* \Omega \;,
\;\;X'\in \cN'\;.
\label{delta12}
\ea
They are closable operators and their closures, also 
denoted by $\check{S}$ and $\check{F}$, admit a polar decomposition 
\be
\check{S} = J\,\Delta^{1/2}\;,\;\;\;{\rm with}\;\;\;\check{S}\check{F} =
\Delta^{-1}\;,\sspace \check{F} \check{S} = \Delta\;,
\label{delta13}
\ee
where $J$ is anti-unitary with respect to the inner product 
on $\cH$ and and $\Delta$ is a positive selfadjoint (in general 
unbounded) operator. The operators $(J,\Delta)$ are the modular 
structures featuring in the Tomita-Takesaki theorem.
Thus, at least in principle, they can be constructed from the 
operators $\check{S},\check{F}$ defined in (\ref{delta12}).

In the present context $s$ in (\ref{delta11}) plays the role
of ${\rm Ad} \check{S}$ in (\ref{delta12}) and the quantum operator 
algebra $\cN_{\beta}(S)$ of section 3.2 plays the role of the von 
Neumann algebra $\cN$ in (\ref{delta12}). Indeed, using the automorphism 
$\sbar$ in (\ref{delta4}) to define ${\it s} = \sbar \circ \sigma$ 
and ${\it f} = \sigma \circ \sbar$, one verifies the following 
properties of $s$ and ${\it f}$: 
Both are anti-linear automorphisms (not anti-automorphisms) of 
$\cM_{\beta}(S)$ or $\cF_{\beta}(S)$. They are involutions 
${\it s}^2 = id,\;{\it f}^2 = id$ and related by 
$ {\it f} = \sigma {\it s} \sigma$. Further 
\begin{subeqnarray}
&& s \circ f = \delta_{-1}\;,\sspace f \circ s = \delta_1\;,\\
&& s(X) \Oket = \sigma(X) \Oket\;,\;\;\; X\in \cN_{\beta}(S)\;,\\
&& (s(X),s(Y)) = \eta^l\,(\sigma(X),\sigma(Y))\;,\;\;\;l\in \Z\;,
\end{subeqnarray}
where $\Oket$ is a T-invariant vector. These properties are 
completely analogous to that of $X \ra \check{S}X\check{S},\;
X \ra \check{F}X\check{F}$ with $\check{S},\check{F}$ defined through 
(\ref{delta12}). Of course the construction principle (\ref{delta13}) 
for the modular structures does not apply here: One is not dealing with 
von Neumann algebras, the state space (\ref{vec9}) in general does not 
carry a positive semi-definite inner product, and the topological 
structure is lacking. Fortunately this is also not needed. The 
modular structures $(j,\delta)$ here are defined {\em directly}
and {\em explicitly} in terms of the $\cT_{\beta}(S)$ generators. 
The automorphisms $s$ and $f$ are derived quantities which can also
be written down explicitly. In summary, the modular structures 
$(j,\delta)$ found here and the `canonical' ones within the 
Tomita-Takesaki theory can be contrasted as follows. 
\begin{itemize}
\item The modular structures $(j,\delta)$ are affiliated with
``quantum operator algebras'' $\cM_{\beta}(S)$ or $\cF_{\beta}(S)$
containing a generalized ``quantum double'' $\cT_{\beta}(S)$
as a subalgebra.
\item They are constructed explicitly in terms of the 
$\cT_{\beta}(S)$ generators, not by means of polar decomposition 
of a closable operator.  
\item The construction does not rely on the existence of positive 
functionals over the algebra $\cM_{\beta}(S)$ or $\cF_{\beta}(S)$. 
The state space $\Sigma$ in (\ref{vec9}) may, but need not, have a positive 
semi-definite inner product. 
\item Topological notions are lacking in the quantum operator algebras
$\cM_{\beta}(S)$ or $\cF_{\beta}(S)$. In particular elements of 
infinite degree are not defined. 
\end{itemize} 
The first three features should probably not be counted as drawbacks,
but the last one certainly calls for improvement. This is because 
local operators in this framework probably are described by elements of 
infinite degree, having among others the property to map solutions
of (\ref{state3}) onto new solutions. In the special case,
where one starts from a positive functional $\ob$ it is plausible that
upon appropriate refinement, the construction of $(j,\delta)$ described
here can be subsumed within the framework of the Tomita-Takesaki 
theory.

So far the algebras $\cM_{\beta}(S)$ and $\cF_{\beta}(S)$ ran 
completely parallel as far as the modular structures $(j,\delta)$ were 
concerned. Of course they are different concerning their role in
the form factor construction: The T-invariant functionals over 
$\cM_{\beta}(S)$ yield solutions of the equivariance equations (I), 
while the functionals over the form factor algebra $\cF_{\beta}(S)$
yield solutions of the combined system (I) and (II). 
One may ask however to what extent they are different with respect to 
the role of the modular structures $(j,\delta)$. We propose
the following answer: For $\cF_{\beta}(S)$ it is possible to recover
the full algebra from the subalgebra $\cW(S)\subset \cN_{\beta}(S)$ and its 
commutant $\cW(S)'\subset \cN_{\beta}(S)'$, while for $\cM_{\beta}(S)$ 
the same is not possible. Though we cannot offer yet a fully fledged 
reconstruction theorem, the following features presumably 
capture the basic ingredients.

We begin by computing the contraction products between the $W$
and the $j W$ generators for generic $\beta$. The first relation (R)
translates into 
\ba
&& T^+(\th)_a^b = -\frac{1}{\lb}C^{bb'}\,
W_a(\th -i\beta +i2\pi)\,\cdot\,j W_{b'}(\th^* +\frac{1}{2}i\beta)\;,
\nonum
&& T^-(\th)_a^b = -\frac{1}{\lb}C^{bb'}\,
j W_{b'}(\th^* +i\frac{1}{2}\beta)\,\cdot\,W_a(\th)\;.
\label{jres1}
\ea
Notice that (\ref{jres1}) is manifestly consistent with the 
(TW) relations just because the $W$-generators commute with the 
$jW$'s. Due to the index contractions the flipped form of (R) does
not give rise to extra relations. However when the S-matrix is  
$2\pi i$-periodic additional relations hold
\begin{subeqnarray}
T^+(\th)_a^b &\, =\,&
-\frac{\dim V}{\lb}C^{bb'}\,
W_a(\th -i\beta)\,\cdot\,j W_{b'}(\th^* +\frac{1}{2}i\beta)\;,
\\ &\, =\,& 
\phantom{-}\frac{\dim V}{\lb} C^{bb'}
j W_{b'}(\th^* +\frac{1}{2}i\beta)\,\cdot\,W_a(\th-i\beta) \\
T^-(\th)_a^b &\, =\,& -\frac{\dim V}{\lb}C^{bb'}\,
j W_{b'}(\th^* +i\frac{1}{2}\beta)\,\cdot\,W_a(\th-i 2\pi) \\
&\, =\,& 
\phantom{-}\frac{\dim V}{\lb}C^{bb'}\,
W_a(\th+2\pi i)\,\cdot\,j W_{b'}(\th^* +i\frac{1}{2}\beta)\;.
\label{jres4}
\end{subeqnarray}
Here (\ref{jres4}b,d) follow from the first relation (R) and (TW).
For (\ref{jres4}b,d) one uses the stronger form of the second  
relation (R), i.e. $W_a(\th -i\pi) \cdot W_b(\th) = -\lb C_{ab}/\dim V$,  
valid when the S-matrix is $2\pi i$-periodic. In either case the derivation 
only works for a non-singular $S_{ab}^{dc}(-i\pi)$; alternatively 
(\ref{jres4}) is clearly consistent with the (TW) relations only if 
$T^{\pm}(\th)$ are $2\pi i$-periodic.  

For $\beta =2\pi$ a computation analogous to that leading to 
(\ref{jres1}) gives 
\ba
T^+(\th)_a^b - T^-(\th)_a^b \is
-\frac{1}{\lb}C^{bb'}\,
W_a(\th)\,\cdot\,j W_{b'}(\th^* +i\pi)
\nonum
\is \phantom{-}\frac{1}{\lb}C^{bb'}\,
j W_{b'}(\th^* +i\pi)\,\cdot\,W_a(\th)\;.
\label{jres5}
\ea
One sees that for generic $\beta$ the $\cT_{\beta}(S)$ generators 
can be recovered from that of $\cW(S)$ and $j\cW(S)$; for $\beta =2\pi$ 
the same holds for their difference. More generally the reconstruction 
theorem envisaged would take two commuting algebras $\cN$ as the starting
point and show that from suitable contraction products between them,
the original algebra $\cF_{\beta}(S)$ can be reconstructed. 
This should facilitate the construction of explicit realizations
of $\cF_{\beta}(S)$. 
\vspace{-1cm}

\newsection{Conclusions}

Since we surveyed the results already in the introduction 
a few comments on the perspective may be appropriate here. 

The study of the representation theory as well as the construction of
realizations of the algebra $\cF_{\beta}(S)$ is an important 
desideratum. The following prospects however seem to make it 
worthwhile to consider. The appearence of a double TTS algebra 
should allow one to make contact to better understood
areas like quantum groups and Bethe Ansatz techniques 
\cite{Fadd,QKZE2,QKZE4,Resh}. The implementation of Lorentz boosts 
with imaginary parameter in terms of the TTS generators can be 
viewed as a QFT counterpart of Baxter's corner transfer matrix 
formalism for integrable models in statistical mechanics 
\cite{Baxter,Davies,Thacker}. Making this relation precise might 
in addition be a route to a Euclidean analogue of modular structures. 

On an algebraic level we expect that the modular structures $(j,\delta)$ 
can be used to give an alternative derivation of Smirnov's 
``local commutativity theorem'' \cite{Smir}. The spin-off of emphasizing
the modular structures underlying it should be to see how the result
generalizes to non-integrable QFTs. 

Finally the algebraic framework described here should prepare the 
ground for generalizations, as mentioned in the introduction.

\vspace{5mm}
{\tt Acknowledgements:} Most of this work was done in Kyoto,
enjoying the hospitality of the Yukawa Institut. I wish to thank
T. Inami, M. Jimbo, H. Konno, V. Korepin, T.. Miwa, M. Pillin 
and R. Sasaki for interesting discussions.

\newpage
\setcounter{section}{0}

\newappendix{Form factor equations}

Here we summarize our conventions for the form factor equations.
We work with a slightly generalized set of form factor equations, 
depending on a real parameter $\beta$. 
For $\beta =2\pi$ they coincide with the form factor equations of 
an integrable massive QFT without bound states. For generic $\beta$ 
one obtains a system of deformed form factor equations, whose solutions 
turn out to define QFTs over some non-commutative space, while leaving 
the S-matrix unchanged \cite{MNunpubl}. Conceptually the solutions to 
both systems of equations are sequences of tensor-valued meromorphic 
functions. The equations are conveniently grouped into two sets (I) and 
(II). The set (I) is a system of equivariance equations characterizing 
the individual members of a sequence, while the set (II) prescribes 
how the solutions of (I) are arranged into sequences.

The input for the (generalized) form factor equations is a given 
two-particle bootstrap S-matrix. To fix our conventions, we repeat 
the defining relations. A matrix-valued meromorphic function
$S_{ab}^{dc}(\th),\;\th\in\C$, is called a two particle 
$S$-matrix if it satisfies the following set of equations. 
First the Yang Baxter equation
\be
S_{ab}^{nm}(\th_{12})S_{nc}^{kp}(\th_{13})S_{mp}^{ji}(\th_{23})
=S_{bc}^{nm}(\th_{23})S_{am}^{pi}(\th_{13})S_{pn}^{kj}(\th_{12})\;,
\label{s1}
\ee
where $\th_{12}=\th_1-\th_2$ etc. Second 
unitarity (\ref{s2}a,b) and crossing invariance (\ref{s2}c)
\begin{subeqnarray}
&&S_{ab}^{mn}(\th)\,S_{nm}^{cd}(-\th)=\delta_a^d\delta_b^c\\
&&S_{an}^{mc}(\th)\,S_{bm}^{nd}(2\pi i-\th)=\delta_a^d\delta_b^c\\
&&S_{ab}^{dc}(\th)=C_{aa'}C^{dd'}\,S_{bd'}^{ca'}(i\pi -\th)\;,
\label{s2}
\end{subeqnarray}
where (\ref{s2}c) together with one of the unitarity conditions
(\ref{s2}a), (\ref{s2}b) implies the other. Further real analyticity and 
bose symmetry
\be
[S_{ab}^{dc}(\th)]^* =S_{ab}^{dc}(-\th^*)\;,\sspace
S_{ab}^{dc}(\th)=S_{ba}^{cd}(\th)\;.
\label{s3}
\ee
Finally the normalization condition
\be
S_{ab}^{dc}(0)=-\delta_a^c\delta_b^d\;,\sspace
\label{s4}
\ee
It is convenient to borrow Penrose's abstract index notation 
from general relativity. That is to say, indices 
$a,b,\ldots$ are not supposed to take numerical values but merely 
indicate the tensorial character of the quantity carrying it.  
Vectors $v^a,\,v^b,\ldots$ for example are elements of (classes of)
abstract modules $V^a,\,V^b,\ldots$ of the same dimensionality $\dim V$.
Covectors $v_a,\,v_b,\ldots$ are elements of the 
dual modules $V_a,\, V_b,\ldots$ and repeated upper and lower
case indices indicate the duality pairing. Indices can be raised
and lowered by means of the (constant, symmetric) `charge conjugation 
matrix' $C_{ab}$ and its inverse $C^{ab}$, satisfying $C_{ad}C^{db}=
\delta_a^b$. The S-matrix is a meromorphic function of $\th$.
Bound state poles, if any, are situated on the imaginary axis in 
the so-called physical strip $0\leq \mbox{Im}\,\th< \pi$. From crossing 
invariance and the normalization (\ref{s4}) one infers that 
$S_{ab}^{dc}(i\pi) = -C_{ab}C^{dc}$ is always regular, in contrast to  
$S_{ab}^{dc}(-i\pi)$ which may be singular. In fact, the relevant 
S-matrices are of one of the following two types: 
\ba
&(a)& S_{ab}^{dc}(\th)\;\;\mbox{is $2\pi i$-periodic}\;,\;\;
\mbox{or else}\nonum
&(b)& S_{ab}^{dc}(-i\pi)\;\;\mbox{is singular}\;.
\label{s5}
\ea

The form factor equations are a system of recursive functional 
equations for tensor-valued meromorphic functions of many variables.
In the algebraic formulation adopted here they arise from the 
$T$-invariant states (\ref{state1}), (\ref{state3}) 
and the relations of the algebra $\cF_{\beta}(S)$. This gives a 
system of functional equations for the matrix elements (\ref{state4}),
whose consistency is ensured by the consistency of the underlying 
algebra. Both for $\beta =2\pi$ and $\beta$ generic two systems of 
equations arise: First a system of equivariance equations%
\footnote{The term is borrowed from \cite{QKZE1}.}
(I) that prescribes their monodromy under the action of an infinite 
discrete group $W_n$ acting on the arguments:
\vspace{4mm}
\eql{I}
\bas \jot5mm
f_A(\th) = L_w(\th)_A^B\,f_B(w\inv \th)\;,\sspace 
L_{w_1w_2}(\th)_A^B = L_{w_1}(\th)_A^C L_{w_2}(w_1\inv\th)_C^B\;.
\eas

Here $L_w(\th)_A^B = L_w(\th_n,\ldots,\th_1)%
_{a_n\ldots a_1}^{b_n\ldots b_1}$ is the matrix representing $w\in W_n$
on the space of $V^{\otimes n}$-valued functions. The group $W_n$ is 
the semidirect product of the permutation group $S_n$ and the 
translation group $\Z^n$ \cite{Cher} and turns out to be generated by 
only two elements, $s_1$ and $\Omega$. Their action on rapidity vectors 
and the corresponding representation matrices are given by
\ba
&\nspace & s_1(\th_n,\ldots,\th_1) = (\th_n,\ldots,\th_3,\th_1,\th_2)\;,
\sspace L_{s_1}(\th)_A^B =\delta_{a_n}^{b_n}\ldots\delta_{a_{3}}^{b_{3}}
S_{a_{2}a_1}^{b_1b_{2}}(\th_{21})\;,\nonum
&\nspace & \Omega(\th_n,\ldots,\th_1) = 
(\th_1 +i\beta,\th_n,\ldots,\th_2)\;,
\;\;\;\;\;\;L_{\Omega}(\th)_A^B =  
\eta\delta_{a_{n-1}}^{b_n}\delta_{a_{n-2}}^{b_{n-1}}\ldots
\delta_{a_1}^{b_2}\delta_{a_n}^{b_1}\;.
\label{f1}
\ea  
$W_n$ can also be considered as a Coxeter group, in which case the 
length of a group element $w \in W_n$ coincides with the power of 
$L_w(\th)$ in the two-particle S-matrix. 
$\Omega^n$ is a central element of $W_n$, which on the functions 
(I) is represented as $\eta^n\delta_A^B$. Explicitly the equivariance 
equations (I) for the elements $s_1,\Omega\in W_n$ are 
\ba 
&& f_{a_n\ldots a_1}(\th_n\ldots \th_1) =
S_{a_2a_1}^{dc}(\th_{21})\,
f_{a_n\ldots a_{n-2}cd}(\th_n,\ldots,\th_3,\th_1,\th_2)\;,\nonum
&& f_{a_n\ldots a_1}(\th_n + i\beta, \th_{n-1},\ldots ,\th_1) =
\eta\,f_{a_{n-1}\ldots a_1 a_n}(\th_{n-1},\ldots,\th_1,\th_n)\;.
\label{f2}
\ea
For later use let us also note the representation matrices for the 
generators $t_j,\;1\leq j\leq n$, of the translation subgroup. Using 
$t_j = s_j\ldots s_{n-1}\Omega s_1\ldots s_{j-1}$, with $s_{i+1} :=
\Omega^{-1} s_i \Omega$, $i=1,\ldots, n-2$, one finds from (I)
and (\ref{f1}) 
\ba
&& t_j(\th_n,\ldots ,\th_1)=
(\th_n,\ldots,\th_j +i\beta,\ldots,\th_1)\;,\nonum
&& L_{t_j}(\th)_A^B =  \eta\,T_{a_j}^c(\th_j|\th_n,\ldots,\th_{j+1})%
^{b_n\ldots b_{j+1}}_{a_n\ldots a_{j+1}}\;
T_c^{b_j}(\th_j-i\beta|\th_{j-1},\ldots,\th_1)%
_{a_{j-1}\ldots a_1}^{b_{j-1}\ldots b_1}\;.
\label{f3}
\ea
Here 
$$
T_{a_n}^{b_n}
(\th_n|\th_{n-1},\ldots,\th_1)_{a_{n-1}\ldots a_1}^{b_{n-1}\ldots b_1} 
= S_{c_{n-1}a_{n-1}}^{b_n b_{n-1}}(\th_{n-1,n})\;
S_{c_{n-2}a_{n-2}}^{c_{n-1} b_{n-2}}(\th_{n-2,n})\;\ldots\;
S_{a_na_1}^{c_2b_1}(\th_{1,n})
$$
is the monodromy matrix; its trace over $a_n=b_n$ yields the
well-known family of commuting operators on $V^{\otimes (n-1)}$. 
The property $L_{t_1\ldots t_n}(\th)_A^B = \eta^n \delta_A^B$
reflects the fact that $\Omega^n = t_1\ldots t_n$ is central.  
The equivariance equations (I) in particular imply that starting with a 
function $f_A(\th)$ analytic in the domain
$Re\,\th_n >\ldots > Re\,\th_1,\;0\leq Im\,\th_i <\beta,\;1\leq i\leq n$,
and is equivariant with respect to $W_n$ the domain of analyticity 
extends to $Re\,\th_{kj}\neq 0,\;\forall k,j$. The equivariance equation
(I) for the translation subgroup are also known as the deformed 
Knizhnik-Zamolodchikov equation (KZE) \cite{FrResh,SmirSG}. 

The second set of form factor equations are residue conditions (II) 
prescribing the residues at the simple poles of the solutions 
of (I). These residues in turn get expressed in terms of solutions 
of (I), but with a lower particle number: $n-2$ in the case of
kinematical poles, considered here. Effectively the equations (II)
thus serve to arrange the solutions of (I) for varying $n$ into
sequences $(f^{(n)})_{n\geq 1}$ such that consecutive (or next to 
consecutive) members of a sequence are related by the residue 
conditions (II). The explicit forms follow from (R) and (\ref{res1}) 
and correspondingly the cases $\beta$ generic and 
$\beta =2\pi$ have to be distinguished. 
\vspace{1mm}
\eqlh{II}
\bas 
&\nspace &\;\mbox{$\beta$ generic:}  \\
&\nspace \nspace & \mbox{Res}\,
f^{(n)}_A(\th_n,\ldots,\th_j + i\pi,\th_j,\ldots ,\th_1)\nonum
&\nspace & \;\;\;\;= -\lb\,C_{a_{j+1}a_j}\,
f^{(n-2)}_{a_n\ldots a_{j+2}a_{j-1}\ldots a_1}
(\th_n,\ldots,\th_{j+2},\th_{j-1},\ldots ,\th_1)\;,
\\ \nonum 
&\nspace & \mbox{Res}\,C^{a_{j+1}a_j}
f^{(n)}_A(\th_n,\ldots,\th_j - i\pi,\th_j,\ldots ,\th_1)
\nonum &\nspace & \;\;\;\;= 
-\lb\,f^{(n-2)}_{a_n\ldots a_{j+2}a_{j-1}\ldots a_1}
(\th_n,\ldots,\th_{j+2},\th_{j-1},\ldots ,\th_1)\;,\nonum
&\nspace& \;\mbox{$\beta=2\pi$:} \\
&\nspace& 
\nspace\mbox{Res}\,f^{(n)}_A(\th_n,\ldots,\th_j+i\pi,\th_j,\ldots,\th_1)
=-\lb\left[\frac{1}{\dim V} 
L_{t_{j+1}}(\th_n,\ldots,\th_j+i\pi,\th_j,\ldots,\th_1)_A^B\;
+ \delta_{A}^{B}\right]\!\times \nonum 
&\nspace & \times 
C_{b_{j+1}b_j}\,
f^{(n-2)}_{b_n\ldots b_{j+2}b_{j-1}\ldots b_1}
(\th_n,\ldots,\th_{j+2},\th_{j-1},\ldots ,\th_1)\;.
\eas
We use the notation ${\rm Res}\,f(\th) = 
i\,{\rm res}_{\th_{j+1} =\th_j \pm i\pi} f(\th)$. The choice 
$\lb =\beta/\pi$ for $\lb$ matches the normalization of the 1-particle states 
${}_b\bra \th_2|\th_1\ket_a = 2\beta \delta_{ba} \delta(\th_{21})$.
Comparing the first and the second equation (II) one sees that for 
generic $\beta$ the residue equations split up into two sets of 
equations, where the right hand sides are independent of $\th_j$.
The poles at $\th_{j+1,j} = \pm i\pi$ and $\th_{j+1,j}=\mp i(\pi-\beta)$ 
are split and merge in the limit $\beta \ra 2\pi$. 

In summary two consecutive members of a sequence of form factors
are related by the following condition: $f^{(n)}_A(\th)$ is regular at
relative rapidities $\th_{j+1,j} =\pm i\beta$, $\beta \neq 2\pi$, and 
has simple poles at relative rapidities $\th_{j+1,j} = \pm i\pi$ with 
the above residues. The equivariance relations (I) lead to further 
poles at relative rapidities 
$\th_{k,j} =\pm i\pi +ip\beta,\;p\in\Z$, whose residues can be
computed from (II). The dependence on $\beta$ in the deformed form
factors will usually be suppressed. When needed to distinguish them
from the undeformed form factors we shall write $(f^{(\beta,n)})_{n\geq 0}$ 
and $(f^{(2\pi,n)})_{n\geq 0}$ for the deformed and undeformed ones, 
respectively. As anticipated by the notation one has 
\be
f^{(2\pi,n)}_A(\th) =\lim_{\beta \ra 2\pi} f_A^{(\beta,n)}(\th)\;.
\label{f4}
\ee
To verify this one has to show that the r.h.s. solves the undeformed 
form factor equations. For the equivariance equations this is obvious.
To see that the residue equations come out correctly, first note that 
the deformed KZE implies that $f^{(\beta,n)}_A(\th)$ also 
has simple poles at $\th_{j+1,j} =-i\pi +i\beta$ with residues
\bas
&\nspace &\mbox{Res}\,f^{(\beta,n)}_A(\th_n,\ldots,\th_j-i\pi
+i\beta,\th_j,\ldots,\th_1) = 
- \frac{\lb}{\dim V} \times \nonum
&\nspace & \times 
L_{t_{j+1}}(\th_n,\ldots,\th_j-i\pi +i\beta,\th_j,\ldots,\th_1)_A^B
\;C_{b_{j+1}b_j}f^{(\beta,n-2)}_{b_n\ldots b_{j+2}b_{j-1}\ldots b_1}
(\th_n,\ldots,\th_{j+2},\th_{j-1},\ldots ,\th_1)\;.
\eas
For $\beta \ra 2\pi$ the poles at 
$\th_{j+1,j}= i\pi$ and $\th_{j+1,j}= i(\beta -\pi)$ merge. They produce 
a simple pole again because by assumption $f^{(\beta,n)}_A(\th)$ does 
not have a pole at $\th_{j+1,j}= i\beta$. In particular this implies
that the residues of the merged poles add up producing the second
equation (II), i.e. the undeformed residue equation.

\newpage 
\newappendix{Quadratic form on $\cF_{\beta}(S)$}
\setcounter{equation}{0}

To define the quadratic form $(\;,\;)_{\omega}$ obeying (\ref{quad1}) 
some preparations are needed. Let $F^{(n)}$ denote the space of functions 
in $n$ real variables that are permutation equivariant and square 
integrable, i.e. $z \in F^{(n)}$ iff
\footnote{Equivariance with respect to the complex conjugate 
representation $s \ra L_s^*(\th)_A^B := [L_s(\th)_A^B]^*$ is imposed 
in order to have the kernel (\ref{Fock2}b) equivariant with respect 
to $L$.} 
\ba
z_A(\th) \is L^*_s(\th)_A^B\,z_B(s^{-1}\th)\,,\sspace \forall s \in S_n\,,
\nonum
\|z\|^2 &\!:=\!& \int \frac{d^n \th}{(2\beta)^n} z_A(\th)^* C^{AB} z_B(\th) 
\;< \infty\,.
\label{Fock1}
\ea
Here $S_n \ni s \ra L_s$ is the representation of the permutation group
inherited from (\ref{f1}). For definiteness let us display the 
representation matrices for the generators $s_i$ 
\be
L_{s_i}(\th)_A^B :=\delta_{a_n}^{b_n}\ldots\delta_{a_{i+2}}^{b_{i+2}}
S_{a_{i+1}a_i}^{b_ib_{i+1}}(\th_{i+1,i})\delta_{a_{i-1}}^{b_{i-1}}\ldots
\delta_{a_1}^{b_1}\;,\;\;\;1\leq i\leq n-1\;,
\label{Lperm}
\ee 
from which all others can be computed. The norm in (\ref{Fock1}) comes  
from an inner product on $F^{(n)}$, which for later use we describe in 
terms of a distributional kernel:
\begin{subeqnarray}
&&\bra z|\tilde{z}\ket = \int 
\frac{d^n\omega}{(2\beta)^n}\frac{d^n\th}{(2\beta)^n}
z_{B'}(\omega)^*C^{B'B}\,{}_{B^T}\bra\omega^T|\,\th\ket_A \,C^{AA'}
\tilde{z}_{A'}(\th)\;,\\ 
&&{}_{B}\bra\omega|\,\th\ket_A 
=\frac{(2\beta)^n}{n!}\sum_{s\in S_n} L_s(\th)_A^C\,C_{CB^T} \;
\delta^{(n)}(\omega^T - s\inv\th)\;,
\label{Fock2} 
\end{subeqnarray}
where $B^T =(b_1,\ldots ,b_n),\;\omega^T =(\omega_1,\ldots,\omega_n)$.
Using $[L_s(s\th)_A^B]^* = C_{AA'}C^{BB'} 
L_{s\inv}(\th^*)_{B'}^{A'}$ the kernel (\ref{Fock2}b) can be checked 
to be hermitian. For equally ordered rapidity vectors 
$\omega_n > \ldots > \omega_1$ and $\th_n > \ldots > \th_1$ only the 
$s =\1$ term survives, and for oppositely ordered ones only the 
$s =\iota$ term. Here $\iota =s_{n-1} \ldots s_2s_1s_2 \ldots s_{n-1}$
is the element of $S_n$ (considered as a Coxeter group) of maximal
length, acting like $\iota(\th_n, \ldots,\th_1) = (\th_1,\ldots,\th_n)$
on rapidity vectors. Averaging (\ref{Fock2}b) with permutation equivariant 
test functions in $F^{(n)}$ one checks $\bra z| z\ket = \|z\|^2$
and hence positive definiteness. Sequences of functions $z^{(n)} \in 
F^{(n)}$ can be used to construct the Fock space representation of 
the ZF-algebra \cite{Lig2,MNunpubl}.   

Permutation equivariant functions can be multiplied by means of the 
following 
\medskip

\noindent {\bf Lemma B1:} Let $g\in F^{(k)}$ and 
$h\in F^{(n-k)}$ be $L^*$-equivariant under the permutation group, i.e. 
\ba
&& g_A(\th) =L^*_s(\th)_A^B\,g_B(s\inv \th) \;,\sspace s\in S_k\;,\nonum
&& h_A(\th) =L^*_s(\th)_A^B\,h_B(s\inv \th) \;,\sspace s\in S_{n-k}\;.
\ea
Let $(I_+,I_-)$ be a partition of $I=(n,\ldots,1)$ into ordered subsets
$I_+=(i_n,\ldots,i_{n-k+1}),\,i_n>\ldots >i_{n-k+1}$ and
$I_-=(i_{n-k},\ldots,i_1),\;i_{n-k}>\ldots >i_1$ and set 
$\th_+=(\th_{i_n},\ldots, \th_{i_{n-k+1}}),\;
\th_-= (\th_{i_{n-k}},\ldots, \th_{i_1})$.   
To each of these ${n \choose k}$ partitions associate an element 
$s(I_+,I_-)\in S_n$ by $s(I_+,I_-)(\th_+,\th_-) =(\th_n,\ldots,\th_1)$.
Then the function $g \circ h \in F^{(n)}$ defined by
\be
(g\circ h)_A(\th) := \sum _{(I_+,I_-)}L^*_{s(I_+,I_-)}(\th)_A^B\,
g_{b_n\ldots b_{n-k+1}}(\th_+)\,h_{b_{n-k} \ldots b_1}(\th_-)
\label{Lprod}
\ee
is $L^*$-equivariant under $S_n$. The multiplication `$\circ$' is 
distributive, associative but non-commutative. The so-defined
algebra of permutation equivariant functions carries a 
$*$-operation $\sigma_{\circ}$ given by 
\be
(\sigma_{\circ}g)_A(\th) =g^*_{A^T}(\th^T)\;,\sspace
\sigma_{\circ}(g \circ h) = \sigma_{\circ}h \circ \sigma_{\circ}g\;.
\label{Lstar}
\ee
The same of course holds for $L$-equivariant functions.
\smallskip

We omit the proof. Observe that in the trivial case where
$S$ is replaced with plus or minus the permutation matrix, the product
(\ref{Lprod}) reduces to the moment multiplication or the wedge 
product, respectively.

For $n,m \geq 0$, let now $G^{(k)}\in {F^{(k)}}^*$, $k= n+m -2 l,\;
l = 0,\ldots,{\rm min}(m,n)$, be a set of L-equivariant 
functions. Use the notation of Lemma B1 with 
$\omega =(\omega_m,\ldots ,\omega_1)$,
$B = (b_m ,\ldots ,b_1)$ and $\th = (\th_n,\ldots ,\th_1)$, 
$A = (a_n,\ldots, a_1)$. Define the following distributional kernel 
\be 
\Gbar_{B A}(\omega|\th) 
= \!\sum_{\forall(J_+,J_-),\forall(I_+,I_-)}\!\!
L_{s(J_-,J_+)}(\omega)_B^C\;
{}_{C_+}\bra \omega_+ | \th_+ \ket _{D_+}\,
G_{C_- D_-}(\omega_- +i\pi,\th_-)\;L_{s(I_+,I_-)}(\th)_A^D\,. 
\label{Lemma2_1}
\ee
Here the sum runs over all partitions of $J=(m,\ldots ,1)$ and 
$I=(n,\ldots,1)$ into pairs of ordered subsets, not just those with
a fixed number of elements $|J_+|$ in $J_+$ or $|I_+|$ in $I_+$ 
as in Lemma B1. Again the indices are not permuted, i.e. 
$D_+= (d_n,\ldots, d_{n-|I_+| +1}),\;D_-= (d_{n-|I_+|},\ldots,d_1)$ etc.. 
Finally ${}_{B}\bra \omega| \th\ket_{A}$ is the kernel (\ref{Fock2}b).
\medskip

{\bf Lemma B2:} The distribution (\ref{Lemma2_1}) has the following 
properties. 

(a) It is $L$-equivariant in both sets of variables $\omega$ 
and $\th$.

(b) It is hermitian in the sense that if there exist $H^{(k)} \in 
{F^{(k)}}^*$ such that $G_C(\th)^* = H_{C^T}({\th^*}^T + i\pi)$
then 
\be
[\Gbar_{B^T A}(\omega^T|\th)]^* =
\Hbar_{A^T B}(\th^T|\omega)\;.
\label{Lemma2_2}
\ee
(c) The number of terms in the sum is (\ref{Lemma2_1}) is  
\be
\#{\rm terms} = { m+n \choose n} =  { m+n \choose m}\;.
\ee

{\em Proof:} (a) Equivariance in $\th$ follows from Lemma B1, which implies
equivariance in $\omega$ once (b) is known. The counting (c) is due to
the constraint $|J_+| = |I_+|$. To verify hermiticity (b) it is convenient  
to make use of the following facts: For any $s \in S_n$ the 
representation matrices $L_s$ satisfy $L^*_s(\iota\th)_{A^T}^{B^T} 
= L_{\iota s\iota}(\th^*)_A^B$. Further for any partition $J =(J_+,J_-)$
of $(n,\ldots ,1)$ there exists a partition $K =(K_+, K_-)$ of 
$(n,\ldots ,1)$ with $|K_+| =|J_+|$ and 
$L_{\iota s(J_-,J_+) \iota}(\th)_A^{D_-D_+} = 
L_{s(K_+,K_-)}(\th)_A^{D_+D_-}$, where $D_{\pm}$ are as above.  
$\Box$

Assignments like 
\be
F^{(n)} \ni x \rra 
W^n[x] =\int\frac{d^n\th}{(2\beta)^n}\,W_{A}(\th) x_{A'}(\th)C^{AA'}\;,
\label{sesq0}
\ee
define distributions over $F^{(n)}$. 
In terms of them we can eventually define the quadratic form 
on $\cF_{\beta}(S)$ associated with a linear form $\ob$. Set
\ba
\nspace (Y,X)_{\omega} &:=& \sum_{\forall(I_+,I_-),\forall(J_+,J_-)}\;
\int\frac{d^m\xi}{(2\beta)^m}\,\frac{d^n\th}{(2\beta)^n}\; 
y_{B'}(\xi)^* C^{B'B} L_{s(J_-,J_+)}(\xi)_B^C \times \nonum
\nspace &&\nspace \times 
{}_{C_+}\bra \xi_+ | \th_+ \ket _{D_+}\;
\ob\left(\sigma(W)_{C_-}(\xi_-)W_{D_-}(\th_-)\right)\;
x_{A'}(\th) C^{A'A} L_{s(I_+,I_-)}(\th)_A^D \;,
\label{sesq1}
\ea
where $X= W^n[x]$ and $Y=W^m[y]$ are two elements of $\cF_{\beta}(S)$. 
Using the $T$-invariance of $\ob$ it is clear that $(\;,\;)_{\omega}$ 
has a unique extension to all of $\cF_{\beta}(S)$. Singularities
encountered for $(\xi_-)_j = (\th_-)_k$ are declared in the sense 
of an $(\xi_-)_j +i\epsilon$, $\epsilon >0$ prescription. Lemma B2 
implies that $(\;,\;)_{\omega}$ is well-defined, contravariant with 
respect to $\sigma$ and hermitian, i.e. satisfies (\ref{quad2}). Further 
it has the announced property (\ref{quad1}). For generic $\ob$ however 
$(\;,\;)_{\omega}$ is not positive semi-definite.



\newpage
\begin{thebibliography}{10}

\bibitem{BBS} O. Babelon, D. Bernard and F.A. Smirnov, 
\CMP{178} (1996) 281 and \CMP{186} (1997) 601.

\bibitem{QKZE4} H. Babujian, M. Karowski and A. Zapetal,
J. Phys. {\bf A 30} (1997) 6425.

\bibitem{Baxter} R.J. Baxter, {\em Exactly Solved Models in Statistical 
Mechanics}, Academic Press, 1982. 

\bibitem{BeLeCl} D. Bernard and A. LeClair, \NP{B399} (1993) 709.

\bibitem{BW} J. Bisognano and E. Wichmann,
J. Math. Phys. {\bf 16} (1975) 985-1007 and 
J. Math. Phys. {\bf 17} (1976) 303-321.

\bibitem{Luk2} V. Brazhnikov and S. Lukyanov,
\NP{B512} (1998) 616.

\bibitem{Cher} I. Cherednik, \CMP{150} (1992) 109.

\bibitem{Davies} B. Davies: Infinite dimensional symmetry
of corner transfer matrices, hep-th/9312141.

\bibitem{DeVega1} H.J. De Vega, H. Eichenherr and  J.M. Maillet,
\CMP{92} (1984) 507.

\bibitem{DeVega2} C. Destri and H.J. De Vega, \NP{B406} (1993) 566.

\bibitem{Ding} J. Ding: Hopf algebra extension of a ZF-algebra 
and its double, q-alg/9612008.

\bibitem{EfLeCl} C. Efthimiou and A. LeClair, \CMP{171} (1995) 531. 

\bibitem{QKZE1} P. Etingof, I. Frenkel and A.A. Kirillov:
Spherical functions on affine Lie groups, hep-th/940747.

\bibitem{EtKa} P. Etingof and D. Kazhdan: Quantization of Lie
Bialgebras V, math.QA/9808121.

\bibitem{Fadd} L. Faddeev, Sov. Sci. Rev. Math. Phys. {\bf C1} (1980)
107. 

\bibitem{FrResh} I.B. Frenkel and N.Yu. Reshetikin, \CMP{146} (1992) 1.

\bibitem{FrRe} E. Frenkel and N.Yu. Reshetikin: Towards deformed
chiral algebras, q-alg/9706023. 

\bibitem{HHW} R. Haag, W. Hugenholtz and M. Winnink,
\CMP{5} (1967) 215.

\bibitem{Iohara} K. Iohara: Bosonic representations of Yangian
double, q-alg/9603033.

\bibitem{JimMiwa1} M. Jimbo and T. Miwa: {\em Algebraic Analysis of 
Solvable Lattice Models}, CBMS Regional Confererence Series in Mathematics,
Vol 85, AMS 1994. 

\bibitem{KarWeisz} M. Karowski and P. Weisz,
\NP{139} (1978) 455.

\bibitem{KLP} S. Khoroshkin, D. Lebedev and S. Pakuliak:
Intertwining operators for the central extension of the Yangian 
double, q-alg/9602030.

\bibitem{KLP2} S. Khoroshkin, D. Lebedev and S. Pakuliak,
Lett. Math. Phys. {\bf 41} (1997) 31. 

\bibitem{Lash} M. Lashkevich, Electronic Journal JHEP10(1997)003; 
hep-th/9704148.

\bibitem{LiZ} B.H. Lian and G.J. Zuckerman: Commutative quantum
operator algebras, J. pure appl. Algebra {\bf 100} (1995) 117.

\bibitem{Lig2} A. Liguori, M. Mintchev and M. Rossi, 
J. Math. Phys. {\bf 38} (1997) 2888. 

\bibitem{Luk1} S. Lukyanov, \CMP{167} (1995) 183.

\bibitem{MNcycl} M. Niedermaier, \CMP{196} (1998) 411.

\bibitem{MNalg}  M. Niedermaier, \NP{B440} (1995) 603; 
Erratum-ibid. {\bf B456} (1995) 755.

\bibitem{MNdeform}  M. Niedermaier, Varying the Unruh temperature 
in integrable QFTs, hep-th/9807041.

\bibitem{MNunpubl}  M. Niedermaier, unpublished.

\bibitem{Ojima} I. Ojima, Ann. Phys. {\bf 137} (1981) 1.

\bibitem{Rehren} K.-H. Rehren, Spin-Statistics and CPT for Solitons,
hep-th/9711085.

\bibitem{Resh} N. Reshetikhin, Lett. Math. Phys. {\bf 26} (1992) 153.

\bibitem{SmirSG} F.A. Smirnov, J. Phys. {\bf A19} (1986) L575.

\bibitem{Smir} Smirnov, F.A.: {\em Form Factors in Completely Integrable
Models of Quantum Field Theory},  World Scientific, 1992.

\bibitem{Smir1} F.A. Smirnov, \NP{B453} (1995) 807.

\bibitem{Smir2} F.A. Smirnov, \CMP{155} (1993) 459.    

\bibitem{Smir3} F.A. Smirnov, Int. J. Mod. Phys. {\bf A7}, Suppl.
1B (1992) 813.

\bibitem{Schroer} B. Schroer, \NP{B499} (1997) 519. 

\bibitem{Thacker} H.B. Thacker, Physica {\bf D18} (1986) 348. 

\bibitem{TT} M. Takesaki, {\em Tomita's theory of modular Hilbert algebras 
and its application}, Lecture Notes in Mathematics, Springer 1970. 

\bibitem{QKZE2} V. Tarasov and A. Varchenko, Amer. Math. Soc. Transl.
(2) {\bf 174} (1996) 235; hep-th/9406060.

\bibitem{Umz} H. Umezawa: {\em Advanced field theory}, American 
Institute of Physics, 1993.

\bibitem{ZZ} A.B. Zamolodchikov and Al.B. Zamolodchikov, 
Ann. Phys. {\bf 120} (1979) 253.

\end{thebibliography}
\end{document}